\documentclass[prb,aps,superscriptaddress,twocolumn,dvips]{revtex4}
\usepackage{feynmp}
\usepackage{amsmath}
\usepackage{amssymb}
\usepackage{epsfig}
\usepackage{graphicx}
\usepackage{subfigure}
\usepackage{hyperref}
\usepackage{bbm}
\usepackage{color}
\usepackage{longtable}
\usepackage{booktabs}
\usepackage{multirow}
\newcommand{\tabincell}[2]{\begin{tabular}{@{}#1@{}}#2\end{tabular}}

\newcommand{\Rmnum}[1]{\expandafter\@slowromancap\romannumeral #1@}
\allowdisplaybreaks[3]
\begin{document}

\title{Effects of the Dirac cone tilt in two-dimensional Dirac semimetal}

\author{Zhao-Kun Yang}
\affiliation{Department of Modern Physics, University of Science and
Technology of China, Hefei, Anhui 230026, China}
\author{Jing-Rong Wang}
\affiliation{Anhui Province Key Laboratory of Condensed Matter
Physics at Extreme Conditions, High Magnetic Field Laboratory of the
Chinese Academy of Science, Hefei, Anhui 230031, China}
\author{Guo-Zhu Liu}
\altaffiliation{Corresponding author: gzliu@ustc.edu.cn}
\affiliation{Department of Modern Physics, University of Science and
Technology of China, Hefei, Anhui 230026, China}

\begin{abstract}
Two-dimensional Dirac semimetal with tilted Dirac cone has recently
attracted increasing interest. Tilt of Dirac cone can be realized in
a number of materials, including deformed graphene, surface state of
topological crystalline insulator, and certain organic compound. We
study how Dirac cone tilting affects the low-energy properties by
presenting a renormalization group analysis of the Coulomb
interaction and quenched disorder. Random scalar potential or random
vector potential along the tilting direction cannot exist on its own
as it always dynamically generates a new type of disorder, which
dominates at low energies and turns the system into a compressible
diffusive metal. Consequently, the fermions acquire a finite
disorder scattering rate. Moreover, the isolated band-touching point
is replaced by a bulk Fermi arc in the Brillouin zone. These results
are not qualitatively changed when the Coulomb interaction is
incorporated. In comparison, random mass and random vector potential
along the non-tilting direction can exist individually, without
generating other types of disorder. They both suppress tilt at low
energies, and do not produce bulk Fermi arc. Upon taking the Coulomb
interaction into account, the system enters into a stable quantum
critical state, in which the fermion field acquires a finite
anomalous dimension but the dynamical exponent $z=1$. These results
indicate that Dirac cone tilt does lead to some qualitatively
different low-energy properties comparing to the untilted system.
\end{abstract}

\maketitle

%%%%%%%%%%%%%%%%%%%%%%%x%%%%%%Main Body%%%%%%%%%%%%%%%%%%%%%%%%%%%%%%%%%%%%%

\section{Introduction}

Semimetal (SM) materials have been studied for more than one decade
\cite{CastroNeto09, Kotov12, Vafek14, Wehling14, Wan11, Weng16,
FangChen16, Yan17, Hasan17, Armitage18}. They exhibit many unusual
properties different from ordinary metals and semiconductors. Among
all the currently known SMs, two-dimensional (2D) Dirac SM (DSM) has
attracted particular interest. Graphene \cite{CastroNeto09, Kotov12}
and the surface state of three-dimensional (3D) topological
insulator \cite{Vafek14, Wehling14} are two sorts of 2D DSM. A
prominent feature of 2D DSM is that, the fermion density of states
(DOS) vanishes at the Fermi level. Thus the Coulomb interaction is
long-ranged and may induce nontrivial quantum corrections to various
observable quantities \cite{Kotov12}. This is quite different from
ordinary metals in which the Coulomb interaction is statically
screened by the finite zero-energy DOS and can be safely neglected
\cite{GiulianiBook, ColemanBook, Shankar94}.

In an intrinsic graphene, the conduction band exhibits an almost
perfect cone shape near the Dirac point. In many realistic DSM
materials, however, the Dirac cone is more or less distorted. It is
known that tilted Dirac cone can be realized in a number of
materials, including properly deformed graphene \cite{Goerbig08,
Choi10}, organic compound $\alpha$-(BEDT-TTF)$_{2}$I$_{3}$
\cite{Goerbig08, Kobayashi07, Hirata16, Hirata17}, 8-$Pmmn$
borophene \cite{Zhou14, Zobolotskiy16, Feng17, Sadhukhan17, Verma17,
Islam17}, and partially hydrogenated graphene \cite{Lu16}. In
addition, the surface of topological crystalline insulators, such as
the (001) surface of SnSe \cite{Tanaka12, Sodemann15}, may also host
a tilted Dirac cone. Recent study \cite{Varyhalov17} suggests that a
tilted Dirac cone appears on the (110) surface of W, which is not
caused by topology but arises from Rashba split bands. To understand
the properties of these materials, it is necessary to study the
physical effects induced by the tilt of Dirac cone. In particular,
we wish to examine whether the tilt leads to qualitatively new
physics. This is the primary motivation of the present work.

For simplicity, we suppose that the Dirac cone is only moderately
tilted such that the Fermi surface is still made out of discrete
zero-dimensional points. In the clean limit, the fermion DOS
vanishes at Fermi level. Isobe and Nagaosa \cite{Isobe12} have
analyzed the impact of long-range Coulomb interaction and revealed
that it is marginally irrelevant. The tilt parameter flows gradually
to zero in the lowest energy limit \cite{Isobe12}. This conclusion
is confirmed by a more recent work \cite{LeeLee18}. It turns out
that tilted DSM exhibits very similar properties to untilted DSM in
the clean limit. However, coupling to disorders may drastically
change the low-energy behaviors of fermions \cite{Papaj18, Zhao18}.

In this paper, we present a renormalization group (RG) study of the
impact of several types of ordinary disorder that commonly exist in
realistic DSM samples, including random scalar potential (RSP),
random vector potential (RVP), and random mass (RM). We are also
interested in the physical effects of the combination of Coulomb
interaction and disorder. It will become clear that the tilt does
lead to interesting properties that are not realized in the untilted
system.

In a untilted 2D DSM, every type of quenched disorder, including
RSP, two components of RVP, and RM, can exist on its own without
generating other types of disorder \cite{Ludwig94, Ostrovsky06,
Evers08, Foster12}. The two components of RVP, abbreviated as
$x$-RVP and $y$-RVP, are equivalent. These features are
significantly changed by the Dirac cone tilt. Without loss of
generality, we suppose the tilt is along $x$-axis. We will show
that, RSP cannot exist alone in the system: it always dynamically
generates a new type of disorder that is originally absent.
Remarkably, the generated disorder is dominant at low energies and
its strength parameter flows to the strong coupling regime. As a
consequence, the system is turned into compressible diffusive metal
(CDM) phase, in which the fermions acquire a finite disorder
scattering rate. Moreover, the original Dirac point is replaced by a
bulk Fermi arc. The $x$-RVP leads to almost the same physical
consequences as RSP. However, $y$-RVP plays an entirely different
role from $x$-RVP, and also from RSP. Interestingly, $y$-RVP can
exist alone in tilted 2D DSM, and it completely suppresses the tilt.
In the low-energy region, $y$-RVP is marginal, leading to power-law
corrections to the energy or temperature dependence of observable
quantities. Analogous to $y$-RVP, RM can also exist independently,
and it also tends to suppress tilt. But RM is marginally irrelevant
in the low-energy region and thus only induces logarithmic-like
corrections to observable quantities.

We see that dirty tilted 2D DSM exhibits distinct physics comparing
to the clean system. It is interesting to investigate how these
results are affected by the Coulomb interaction. In the cases of RSP
and $x$-RVP, the Coulomb interaction is much less important, and the
system is still turned into CDM that exhibits bulk Fermi arc. If
there is no tilt, strong Coulomb interaction dominates RSP at low
energies and protects the SM phase, whereas the interplay of Coulomb
interaction and $x$-RVP leads to a stable quantum critical state
featuring vanishing DOS at Fermi level. Obviously, this difference
is caused by the tilt. When the Coulomb interaction cooperates with
$y$-RVP or RM, the tilt flows to zero in the lowest energy limit,
and the system enters into a stable quantum critical state, where
the fermion field acquires a finite anomalous dimension but the
dynamical exponent is $z=1$. To gain a better understanding of this
critical state, we calculate several observable quantities, such as
fermion DOS, specific heat, and compressibility, and analyze their
energy or temperature dependence.

The rest of the paper will be organized as follows. The effective
model for tilted 2D Dirac fermions is described in
Sec.~\ref{Sec:Model}. The dynamical generation of new type of
disorder is discussed in this section. The coupled RG equations for
various model parameters are presented in
Sec.~\ref{Sec:RGEquations}. The numerical results for the RG
equations and their physical implications are analyzed in
Sec.~\ref{Sec:NumResutls}. In Sec.~\ref{Sec:Summary}, the main
results obtained in this work are summarized and compared to
previous works. Detailed RG calculations are presented in
Appendix~\ref{App:DerivationRG}. Observable quantities are
calculated in Appendixes~\ref{App:ObservableQuantities} and
\ref{App:ObservableQuantitiesSQCS}. In
Appendix~\ref{App:RGResultsDisgardingImportantTerm}, we present the
RG results obtained by omitting the dynamically generated
fermion-disorder coupling.

\section{Effective Model \label{Sec:Model}}

We suppose that the Dirac cone is tilted along $x$-axis. The free
Hamiltonian for 2D tilted Dirac fermions is
\begin{eqnarray}
H = \int_{\mathbf{x}}\psi^{\dag}(\mathbf{x})\left[-v_{x}
\left(t\sigma_{0}+\sigma_{x}\right)i\partial_{x}-v_{y}\sigma_{y}
i\partial_{y}\right]\psi(\mathbf{x}),
\end{eqnarray}
where $\int_{\mathbf{x}}\equiv \int d^{2}\mathbf{x}$ and $\psi$ is a
two-component spinor. There are $N$ copies of $\psi$, where $N$ can
be identified as the fermion flavor. For graphene, the two
components of $\psi$ correspond to the two sublattices $A$ and $B$.
The physical fermion flavor \cite{Kotov12, Vafek07, Sheehy07} is $N
= 4$, which originates from the two valleys ($K$, $K'$) and the two
spins ($\uparrow$, $\downarrow$). $\sigma_{x}$ and $\sigma_{y}$ are
two Pauli matrices, and $\sigma_{0}$ is identity matrix. The energy
dispersion of fermions is defined as
\begin{eqnarray}
E_{\pm} = t v_{x}k_{x} \pm \sqrt{v_{x}^{2}k_{x}^{2} +
v_{y}^{2}k_{y}^{2}},
\end{eqnarray}
where $v_{x}$ and $v_{x}$ are two velocities. The tilt parameter is
$t$ with $t=0$ corresponding to untilted Dirac cone. For $|t|<1$,
the system is identified as type-I DSM, in which the Fermi surface
is made out of discrete band-touching points. For $|t|>1$, the
system corresponds to type-II DSM, whose Fermi surface is a finite
open line \cite{LeeLee18, Soluyanov15, Isobe16TyepII, Huang17}. The
fermion DOS vanishes at the Fermi level in type-I DSM, but takes a
finite value in type-II DSM. Physically, $t<0$ or $t>0$ are
equivalent, thus we only consider the case of $0<t<1$ hereafter.

The Hamiltonian for Coulomb interaction is
\begin{eqnarray}
H_{C} = \frac{1}{4\pi}\int d^2\mathbf{x} d^2\mathbf{x}'
\rho(\mathbf{x}) \frac{e^{2}}{\epsilon\left|\mathbf{x} -
\mathbf{x}'\right|}\rho(\mathbf{x}'),
\end{eqnarray}
where $\rho(\mathbf{x}) = \psi^{\dag}(\mathbf{x})\psi(\mathbf{x})$
is fermion density, $e$ electric charge and $\epsilon$ dielectric
constant. The role of Coulomb interaction depends on the fermion
dispersion and dimension \cite{Gonzalez99, Son07, Barnes14,
Hofmann14, Sharma16, Goswami11, Hosur12, Gonzalez14, Hofmann15,
Throckmorton15, Sharma18, Yang14, Abrikosov72, Abrikosov71,
Abrikosov74, Moon13, Herbut14, Janssen15, Dumitrescu15, Janssen16,
Janssen17, Huh16, Isobe16A, Cho16, Isobe16B, Lai15, Jian15, Zhang17,
Isobe12, Detassis17, LeeLee18}. RG studies found that Coulomb
interaction is marginally irrelevant in 2D DSM and induces
logarithmic-like corrections to the energy, momenta, or temperature
dependence of observable quantities \cite{Kotov12, Gonzalez99,
Son07, Barnes14, Hofmann14}. For instance, the fermion velocity is
predicted  to acquire a logarithmic-like momenta-dependence
\cite{Kotov12, Gonzalez99, Son07, Barnes14, Hofmann14}, consistent
with experiments \cite{Elias11, Siegel11, Yu13, Miao13, Faugeras15}.

The free fermion propagator is
\begin{eqnarray}
G_{0}\left(i\omega,\mathbf{k}\right) = \frac{1}{i\omega \sigma_{0} -
\left(t\sigma_{0}+\sigma_{x}\right)v_{x}k_{x}-v_{y}k_{y}\sigma_{y}}.
\label{Eq:FermionPropagator}
\end{eqnarray}
The bare Coulomb interaction in momentum space is
\begin{eqnarray}
D_{0}(\mathbf{q}) = \frac{2\pi e^{2}}{\epsilon|\mathbf{q}|}.
\end{eqnarray}
One can carry out perturbative expansion in powers of small Coulomb
strength parameter and compute the fermion self-energy by using the
bare function $D_{0}(\mathbf{q})$. However, this expansion scheme is
valid only for weak coupling. Here, we choose to adopt the $1/N$
expansion scheme \cite{Son07, Hofmann14}, which works well in both
the weak and strong coupling regimes \cite{Hofmann14}. In order to
implement $1/N$ expansion, we employ the dressed Coulomb interaction
\begin{eqnarray}
D(\mathbf{q}) = \frac{1}{D_{0}^{-1}(\mathbf{q})+\Pi(\mathbf{q})},
\label{Eq:DressedCoulomb}
\end{eqnarray}
where $\Pi(\mathbf{q})$ is the static polarization function. At the
one-loop order, it is straightforward to find that
\begin{eqnarray}
\Pi(\mathbf{q}) &=& -N\int \frac{d\omega}{2\pi}
\frac{d^2\mathbf{k}}{(2\pi)^{2}} \mathrm{Tr}
\left[G_{0}(i\omega,\mathbf{k})G_{0}(i\omega,\mathbf{k}+\mathbf{q})\right]
\nonumber
\\
&=& \frac{N}{16v_{x}v_{y}}\frac{v_{x}^{2}q_{x}^{2}+v_{y}^{2}
q_{y}^{2}}{\sqrt{v_{x}^{2}q_{x}^{2}+v_{y}^{2}q_{y}^{2}-t^{2}
v_{x}^{2}q_{x}^{2}}}.
\end{eqnarray}
The dressed interaction propagator Eq.~(\ref{Eq:DressedCoulomb})
contains a factor of $N$ in the denominator, thus the Feynman
diagrams with higher powers of Eq.~(\ref{Eq:DressedCoulomb}) will be
suppressed by large $N$. The $1/N$ expansion scheme is applicable to
both weak and strong Coulomb interactions \cite{Hofmann14}.

The disorder effects can be incorporated by introducing the
following fermion-disorder coupling term
\begin{eqnarray}
S^{\mathrm{dis}} = \sum_{j=0}^{3}\int d\tau d^2\mathbf{x}
V_{j}\psi^{\dag}\Gamma_{j}\psi.
\end{eqnarray}
The random potential $V_{j}$ is assumed to be a Gaussian white
noise, satisfying the relations $\langle V_{j}(\mathbf{x})\rangle =
0$ and $\langle V_{j}(\mathbf{x})V_{j}(\mathbf{x}')\rangle =
\Delta_{j}\delta^{2}(\mathbf{x}-\mathbf{x}')$. Averaging over the
random field by applying the replica trick, we re-cast the above
action into
\begin{eqnarray}
S^{\mathrm{dis}} = \sum_{j=0}^{3}\frac{\Delta_{j}}{2}\int d\tau
d\tau' d^{2}\mathbf{x} \left(\psi_{a}^{\dag}\Gamma_{j}
\psi_{a}\right)_{\tau} \left(\psi_{b}^{\dag}
\Gamma_{j}\psi_{b}\right)_{\tau'}, \label{Eq:Sdis}
\end{eqnarray}
where $a,b=1,2,..., n$ are replica indices. At the end of
calculation, the limit $n\rightarrow 0$ will be taken. The random
fields is classified by the definition of the matrix $\Gamma_{j}$.
RSP is defined by $\Gamma_{0} = \sigma_{0}$. Matrices $\Gamma_{1,2}
= \sigma_{x,y}$ correspond to the two components of RVP. For
$\Gamma_{j} = \sigma_{3}$, disorder behaves as RM. $\Delta_{j}$ with
$j = 0,1,2,3$ are the fermion-disorder coupling coefficients.

In some 2D DSM materials, such as graphene, RSP can be generated by
local defects, neutral impurity atoms, or neutral absorbed atoms
\cite{Peres10, Mucciolo10}. RM is usually produced by the random
configurations of the substrates \cite{Champel10, Kusminskiy11}. RVP
is induced by ripples under proper conditions \cite{Meyer07,
Herbut08}. These disorders fall into four types: RSP, $x$-RVP,
$y$-RVP, and RM. We suppose that initially the system contains only
one type of disorder, and then study each of the four types
separately.

The Coulomb interaction and disorder scattering are important in
almost all SM materials. Their interplay may result in a variety of
striking consequences, such as non-Fermi-liquid (NFL) behavior and
some quantum phase transitions \cite{Ye98, Ye99, Stauber05,
Herbut08, Vafek08, Foster08, WangLiu14, Moon14, Zhao16, Gonzalez17,
Nandkishore17, YuXuanWang17, WangLiuZhang17MWSM, Mandal17}. If
disorder dominates over Coulomb interaction, the SM state might be
turned into a CDM state. Under certain circumstances, the Coulomb
interaction can suppress disorder at low energies and the system is
still in stable SM phase \cite{Kotov12, Stauber05, Herbut08,
Vafek08, WangLiu14, WangLiuZhang17MWSM}. To unveil the true ground
state, one needs to carefully study the Coulomb interaction and
fermion-disorder coupling in a self-consistent way. The RG method
provides an ideal tool for this study.

We will apply the RG method to demonstrate that, intriguing new
features emerge when the Dirac cone is tilted. As long as $t \neq
0$, the fermion self-energy induced by disorder scattering contains
a term $\sim t i\omega \sigma_{x}$ that is originally absent in the
free fermion propagator. This term cannot be simply ignored. To
properly account for such a dynamically generated kinetic term, we
add by hand a term $\lambda i\omega\sigma_{x}$ in the free fermion
action, and then examine how $\lambda$ flows under RG
transformations. This method has been used by Sikkenk and Fritz
\cite{Sikkenk17} to study the disorder effects in 3D tilted Weyl SM
(WSM).

According to the above analysis, we need to start from the following
free propagator
\begin{eqnarray}
G_{0}^{-1}(i\omega,\mathbf{k}) &=& i\omega\sigma_{0} - (t \sigma_{0}
+ \sigma_{x})v_{x}k_{x} - \lambda i\omega\sigma_{x}\nonumber
\\
&& - v_{y}k_{y}\sigma_{y}. \label{Eq:FermionPropagatorNew}
\end{eqnarray}
The corresponding dispersion is determined by
\begin{eqnarray}
\left|\left|E-tv_{x}k_{x}-\lambda E\sigma_{x} -
v_{x}k_{x}\sigma_{x}-v_{y}k_{y}\sigma_{y}\right|\right| = 0.
\end{eqnarray}
Solving this equation, we obtain
\begin{eqnarray}
E_{\pm}(\mathbf{k})=\frac{t+\lambda}{1-\lambda^{2}}v_{x}k_{x}\pm
\sqrt{\frac{\left(1+t\lambda\right)^{2}}{\left(1 -
\lambda^{2}\right)^{2}} v_{x}^{2}k_{x}^{2} + \frac{v_{y}^{2}
k_{y}^{2}}{1-\lambda^{2}}}.
\end{eqnarray}
Define three new effective parameters
\begin{eqnarray}
v_{x}^{\mathrm{eff}} &=& \frac{1+t\lambda}{1-\lambda^{2}}v_{x}, \\
v_{y}^{\mathrm{eff}} &=& \frac{1}{\sqrt{1-\lambda^{2}}}v_{y},
\\
t^{\mathrm{eff}} &=& \frac{t+\lambda}{1+t\lambda}.
\end{eqnarray}
Then $E_{\pm}$ can be further written as
\begin{eqnarray}
E_{\pm}(\mathbf{k}) = t^{\mathrm{eff}}v_{x}^{\mathrm{eff}}k_{x}
\pm\sqrt{\left(v_{x}^{\mathrm{eff}}\right)^{2}k_{x}^{2} +
\left(v_{y}^{\mathrm{eff}}\right)^{2}k_{y}^{2}}.
\end{eqnarray}
It is more convenient to characterize the Dirac cone tilting by the
new parameter $t^{\mathrm{eff}}$. In the RG analysis, the initial
value of $\lambda$ is taken as $\lambda_{0}=0$.

A unique consequence of Dirac cone tilting is that the interaction
between fermions and RSP (or $x$-RVP) can generate a new coupling
term
\begin{eqnarray}
(\psi^{\dag}\sigma_{0}\psi) (\psi^{\dag}\sigma_{x}\psi)
\label{Eq:DisorderGenerateTerm}
\end{eqnarray}
in the process of RG calculations. This term connects two different
bilinear fermion operators $\psi^{\dag}\sigma_{0}\psi$ and
$\psi^{\dag}\sigma_{x}\psi$. Such a mixed term is also absent in the
original action Eq.~(\ref{Eq:Sdis}). We believe that this new
coupling should not be naively discarded. An important reason is
that, this coupling is generated only when the tilt parameter $t$ is
nonzero and thus may be an intrinsic feature of tilted DSM.
Moreover, the new coupling has feedback effects on the dynamics of
Dirac fermions: it results in vertex corrections to the original
fermion-disorder coupling terms given in Eq.~(\ref{Eq:Sdis}). To
handle the generated disorder, we add to the original action the
following coupling term
\begin{eqnarray}
S^{\mathrm{dis}}_{-} = \frac{\Delta_{-}}{2}\int d\tau d\tau'
d^{2}\mathbf{x}
\left(\psi_{a}^{\dag}\Gamma_{-}\psi_{a}\right)_{\tau}
\left(\psi_{b}^{\dag}\Gamma_{-}\psi_{b}\right)_{\tau'},
\label{Eq:SdisMinus}
\end{eqnarray}
where the matrix $\Gamma_{-}=\frac{1}{2}(\sigma_{0}-\sigma_{x})$. It
is easy to verify that the dynamically generated term
$(\psi^{\dag}\sigma_{0}\psi) \left(\psi^{\dag}\sigma_{x}\psi\right)$
can be decomposed into three fermion-disorder coupling terms, namely
$\left(\psi^{\dag}\Gamma_{-}\psi\right)
\left(\psi^{\dag}\Gamma_{-}\psi\right)$,
$\left(\psi^{\dag}\sigma_{0}\psi\right)
\left(\psi^{\dag}\sigma_{0}\psi\right)$, and
$\left(\psi^{\dag}\sigma_{x}\psi\right)
\left(\psi^{\dag}\sigma_{x}\psi\right)$. We will perform RG
calculations based on the total action that contains
Eq.~(\ref{Eq:SdisMinus}).

\section{The RG equations \label{Sec:RGEquations}}

In actual materials, the Coulomb interaction and the quenched
disorders may exist in various kinds of combination. We consider the
most generic case in which the Coulomb interaction and all the
possible disorders coexist in the same tilted DSM system. To deal
with such a complicated model, we will perform a systematic RG
analysis and derive the flow equations for all the involved model
parameters, including the fermion velocities, the velocity ratio,
the Coulomb interaction strength, the tilt parameter, and the
disorder strength parameter. To ensure the validity of RG analysis,
we suppose that the fermion-disorder coupling is weak.

The calculational details of the RG analysis are presented in
Appendix~\ref{App:DerivationRG}. Here, we only list the complete set
of coupled RG equations. Various limiting cases can be easily
obtained from these equations. In particular, the coupled RG
equations are given by
\begin{widetext}
\begin{eqnarray}
\frac{\partial v_{x}}{\partial\ell}&=&\left[M^{\mathrm{dis}}_{A}
+ \mathcal{H}_{1}(\alpha)\right]v_{x}, \label{Eq:RGEqvx}
\\
\frac{\partial v_{y}}{\partial\ell} &=& \left[M^{\mathrm{dis}}_{A} +
\mathcal{H}_{2}(\alpha)\right]v_{y}, \label{Eq:RGEqvy}
\\
\frac{\partial(v_{y}/v_{x})}{\partial\ell} &=&
\left[\mathcal{H}_{2}(\alpha) - \mathcal{H}_{1}(\alpha)\right]
\frac{v_{y}}{v_{x}}, \label{Eq:RGEqvRatio}
\\
\frac{d(tv_{x})}{d\ell} &=& \left[M^{\mathrm{dis}}_{A}t +
\mathcal{H}_{3}(\alpha)\right]v_{x}, \label{Eq:RGEqwvx}
\\
\frac{dt}{d\ell} &=& \mathcal{H}_{3}(\alpha) -
\mathcal{H}_{1}(\alpha)t, \label{Eq:RGEqw}
\\
\frac{d\lambda}{d\ell} &=& M^{\mathrm{dis}}_{A} \lambda +
M^{\mathrm{dis}}_{B}, \label{Eq:RGEqLambda}
\\
\frac{d\alpha}{d\ell} &=& -\left[M^{\mathrm{dis}}_{A}+\frac{1}{2}
\left(\mathcal{H}_{1}(\alpha) + \mathcal{H}_{2}(\alpha)
\right)\right]\alpha, \label{Eq:RGEqAlpha} \\
\frac{d\Delta_{0}}{d\ell} &=& \left(1-t^{2}\right)^{-3/2}
\bigg[\left(2-t\right)\Delta_{0}^{2}+2\left(1-t\right)
\Delta_{0}\Delta_{1}+\left(2+t\right)\Delta_{0}
\left(\Delta_{2}+\Delta_{3}\right)-t\Delta_{1}^{2} -
t\Delta_{1}\Delta_{2}\nonumber \\
&& +\left(4-t-4t^2\right)\Delta_{1}\Delta_{3} + 4\left(1+t\right)
\Delta_{2}\Delta_{3}+\left(\frac{3}{4}+\frac{t}{2}-\frac{t^{2}}{4}
\right) \Delta_{0}\Delta_{-}+\left(\frac{1}{4}-\frac{t}{2} -
\frac{3t^{2}}{4}\right)\Delta_{1}\Delta_{-} \nonumber
\\
&& +\frac{\left(1+t\right)^{2}}{4}\left(\Delta_{2}+\Delta_{3}\right)
\Delta_{-}\bigg]+\mathcal{H}_{4}(\alpha),\label{Eq:RGEqDelta0}
\\
\frac{d\Delta_{1}}{d\ell} &=& \left(1-t^{2}\right)^{-3/2}
\bigg[-t\Delta_{0}^{2}-2t\left(1-t\right)\Delta_{0}\Delta_{1} +
t\Delta_{0}\Delta_{2} + \left(4+t-4t^{2}\right)\Delta_{0}\Delta_{3}
- t\left(1-2t\right)\Delta_{1}^{2} \nonumber \\
&&-t\left(1+2t\right)\Delta_{1}\left(\Delta_{2}+\Delta_{3}\right)
+4t\left(1+t\right)\Delta_{2}\Delta_{3}-\left(\frac{3}{4} +
\frac{t}{2}-\frac{t^{2}}{4}\right)\Delta_{0}\Delta_{-} -
\left(\frac{1}{4}-\frac{t}{2}-\frac{3t^{2}}{4}\right)
\Delta_{1}\Delta_{-} \nonumber \\
&& -\frac{\left(1+t\right)^{2}}{4} \left(\Delta_{2} +
\Delta_{3}\right)\Delta_{-}\bigg]+\mathcal{H}_{5}(\alpha),
\label{Eq:RGEqDelta1}
\\
\frac{d\Delta_{2}}{d\ell} &=& \left(1-t^{2}\right)^{-3/2}
\left[4\left(\Delta_{0}+t^{2}\Delta_{1}\right)\Delta_{3} +
\left(1+t\right)^{2}\Delta_{3} \Delta_{-}\right] + \mathcal{H}_{6}(\alpha),
\label{Eq:RGEqDelta2}
\\
\frac{d\Delta_{3}}{d\ell} &=& \left(1-t^{2}\right)^{-3/2}
\bigg[4\left(1-t^{2}\right)\Delta_{0}\Delta_{1} + 4\Delta_{0}
\Delta_{2} + 4t^{2}\Delta_{1}\Delta_{2}-2\left(1-t^{2}\right)
\left(\Delta_{0}-\Delta_{1}-\Delta_{2}+\Delta_{3}\right)\Delta_{3}
\nonumber \\
&&+\left(1-t^{2}\right)\left(\Delta_{0}+\Delta_{1}\right)\Delta_{-}
+ \left(1+t\right)^{2}\Delta_{2}
\Delta_{-}\bigg]+\mathcal{H}_{7}(\alpha), \label{Eq:RGEqDelta3}
\\
\frac{d\Delta_{-}}{d\ell} &=& \left(1-t^{2}\right)^{-3/2}
\bigg[4t\left(\Delta_{0}^{2}+2\Delta_{0}\Delta_{1} -
\Delta_{0}\Delta_{2}-\Delta_{0}\Delta_{3} +
\Delta_{1}^{2}+\Delta_{1}\Delta_{2}+\Delta_{1}\Delta_{3} -
2\Delta_{2}\Delta_{3}\right) \nonumber \\
&& +\left(3+4t+t^{2}\right)\Delta_{0}\Delta_{-} + \left(1+4t +
3t^{2}\right)\Delta_{1}\Delta_{-} + \left(1-t^{2}\right)
\left(\Delta_{2} + 5\Delta_{3}\right)\Delta_{-} +
\left(1+t\right)^{2} \Delta_{-}^{2}\bigg] \nonumber \\
&& +\mathcal{H}_{8}(\alpha).\label{Eq:RGEqDeltaPlus}
\end{eqnarray}
\end{widetext}
Here, we define the following parameters and functions:
\begin{eqnarray}
M^{\mathrm{dis}}_{A} &=& -\frac{1+t\lambda}{\left(1-t^{2}
\right)^{3/2}}\Big[\sum_{j=0}^{3}\Delta_{j}+\left(1+t\right)
\frac{\Delta_{-}}{2}\Big], \\
M^{\mathrm{dis}}_{B} &=& \frac{1+t\lambda}{\left(1-t^{2}
\right)^{3/2}} \Big[t\left(\Delta_{0}+\Delta_{1}-\Delta_{2} -
\Delta_{3}\right)\nonumber \\
&&+\left(1+t\right)\frac{\Delta_{-}}{2}\Big],\\
\mathcal{H}_{1}(\alpha)&=&\left(1+t\lambda\right)f_{A}(\alpha),
\\
\mathcal{H}_{2}(\alpha)&=&\left(1+t\lambda\right)^{2}f_{B}(\alpha),
\\
\mathcal{H}_{3}(\alpha)&=&\lambda\left(1+t\lambda\right)f_{A}(\alpha),
\\
\mathcal{H}_{4}(\alpha)&=&-\left(1+t\lambda-\lambda-2\lambda^{2}\right)
f_{A}(\alpha) \Delta_{0}\nonumber
\\
&&-\left(1+t\lambda\right)^{2}f_{B}(\alpha)\Delta_{0} + \lambda
f_{A}(\alpha)\Delta_{1}\nonumber
\\
&&-\frac{1}{4}\left(1-\lambda^{2}\right)f_{A}(\alpha)\Delta_{-},
\\
\mathcal{H}_{5}(\alpha)&=&\lambda f_{A}(\alpha)\Delta_{0} +
\left(1+\lambda-t\lambda\right)f_{A}(\alpha)\Delta_{1}\nonumber
\\
&&-\left(1+t\lambda\right)^{2}f_{B}(\alpha)\Delta_{1}\nonumber
\\
&&+\frac{1}{4}\left(1-\lambda^{2}\right)f_{A}(\alpha)\Delta_{-},
\\
\mathcal{H}_{6}(\alpha)&=&\left(1+t\lambda\right)\left[-f_{A}(\alpha)
+ \left(1+t\lambda\right)f_{B}(\alpha)\right]\Delta_{2}, \\
\mathcal{H}_{7}(\alpha) &=& \left[\left(1-2\lambda^{2} -
t\lambda\right)f_{A}(\alpha) \right. \nonumber
\\
&&\left.+\left(1+t\lambda\right)^{2}f_{B}(\alpha)\right]\Delta_{3},
\\
\mathcal{H}_{8}(\alpha) &=& -4\lambda f_{A}(\alpha)
\left(\Delta_{0}+\Delta_{1}\right) + \left(-2\lambda +
\lambda^{2}-t\lambda\right)\nonumber
\\
&&\times f_{A}(\alpha)\Delta_{-} - \left(1+t\lambda\right)^{2}
f_{B}(\alpha)\Delta_{-}.
\end{eqnarray}
The parameter that measures the effective Coulomb interaction
strength can be defined as
\begin{eqnarray}
\alpha=\frac{e^{2}}{\epsilon\sqrt{v_{x}v_{y}}}.
\end{eqnarray}
The functions $f_{A}$, $f_{B}$, and $f_{C}$ are given by
\begin{eqnarray}
f_{A}(\alpha)&=&\frac{\alpha}{\pi} \int_{0}^{\pi/2} d\theta
\sin^{2}\theta\mathcal{G}(\theta,\alpha), \label{Eq:fA}
\\
f_{B}(\alpha)&=&\frac{\alpha}{\pi} \int_{0}^{\pi/2}d\theta
\cos^{2}\theta \mathcal{G}(\theta,\alpha), \label{Eq:fB}
\\
f_{C}(\alpha)&=&\frac{\alpha}{\pi} \int_{0}^{\pi/2}d\theta
\mathcal{G}(\theta,\alpha), \label{Eq:fC}
\end{eqnarray}
where
\begin{eqnarray}
\mathcal{G}^{-1}(\theta,\alpha) &=&
\left[\left(1+t\lambda\right)^{2} \cos^{2}\theta +
\left(1-\lambda^{2}\right) \sin^{2}\theta
\right]^{3/2} \nonumber \\
&& \times \left[\sqrt{\frac{v_{y}}{v_{x}}\cos^{2}\theta +
\frac{v_{x}}{v_{y}}\sin^{2}\theta}\right.\nonumber
\\
&&\left.+\frac{\pi N\alpha}{4}\frac{1}{\sqrt{1-t^{2}
\cos^{2}\theta}}\right]. \label{Eq:ParameterG}
\end{eqnarray}
In the RG analysis, we have made the replacement
\begin{eqnarray}
\frac{\Delta_{j}}{2\pi v_{x}v_{y}}\rightarrow\Delta_{j}.
\end{eqnarray}
The RG equations are derived by integrating the fast modes defined
within the momentum shell $b\Lambda < E_{\mathbf{k}} < \Lambda$,
where $E_{\mathbf{k}} = tv_{x}k_{x} + \sqrt{v_{x}^{2}k_{x}^{2} +
v_{y}^{2}k_{y}^{2}}$ and $b = e^{-\ell}$ with $\ell$ being a running
parameter. The lowest energy limit is approached as $\ell
\rightarrow \infty$.

\section{Physical interpretation of numerical solutions \label{Sec:NumResutls}}

The RG equations are analyzed in this section. Firstly, we consider
the Coulomb interaction in clean limit. Secondly, we study the
disorder effects. Finally we analyze the interplay of Coulomb
interaction and disorder.

\subsection{Pure Coulomb interaction \label{SubSec:NumResutlsCoulomb}}

\begin{figure}[htbp]
\center
\includegraphics[width=3.35in]{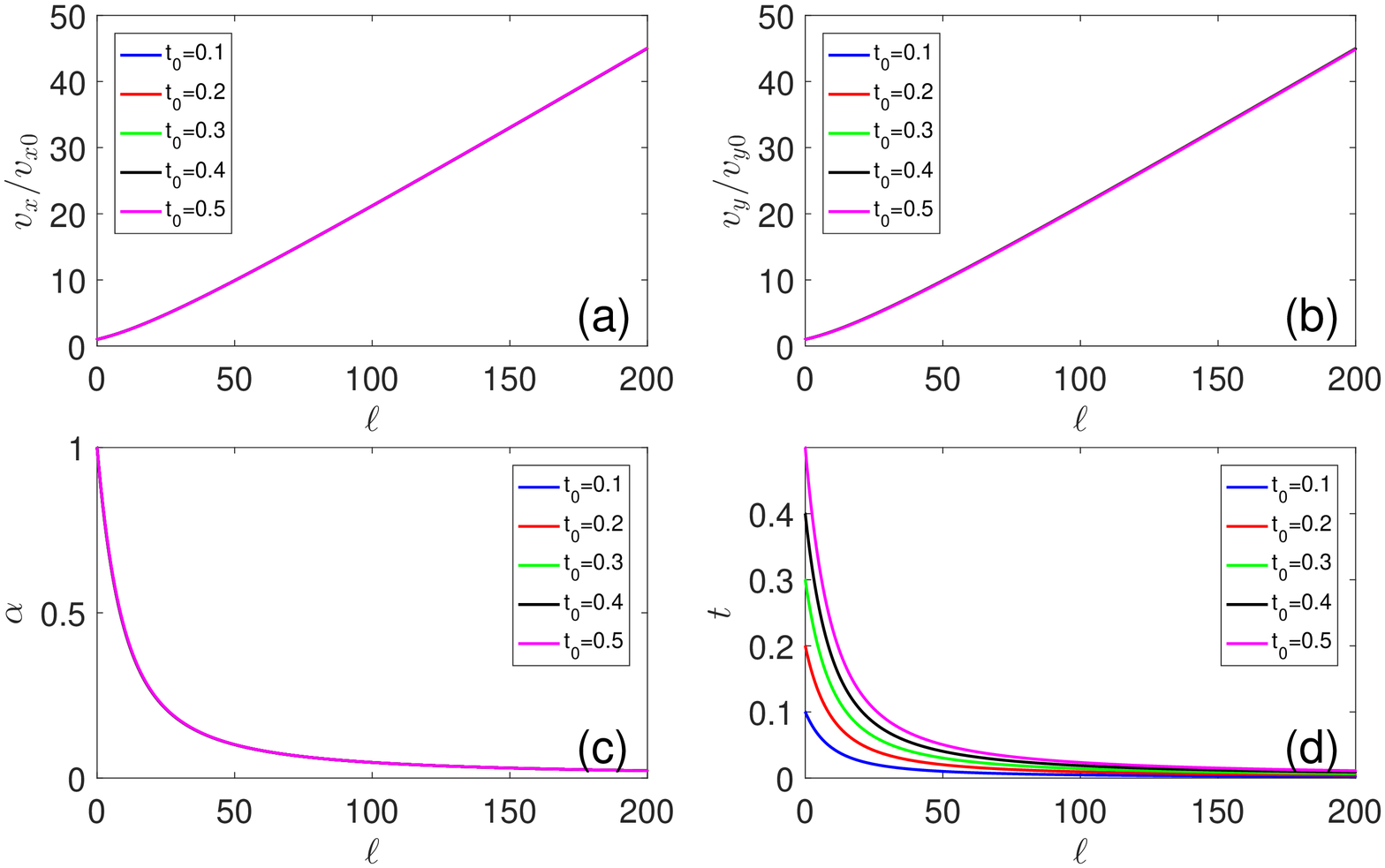}
\caption{Flows of $v_{y}$, $v_{y}$, $\alpha$, and $t$ due to Coulomb
interaction. Here, we assume $v_{y0}/v_{x0}=1$ and $\alpha_{0}=0.5$.
The flavor $N = 4$. \label{Fig:VRGClean}}
\end{figure}

The influence of Coulomb interaction in the clean limit has already
been studied by Isobe and Nagaosa \cite{Isobe12}, and Lee and
Lee\cite{LeeLee18}. Here, we will recover their results. After
removing all the disorders, the coupled RG equations are simplified
to
\begin{eqnarray}
\frac{d v_{x}}{d\ell} &=& f_{A}(\alpha)v_{x},
\\
\frac{d v_{y}}{d \ell} &=& f_{B}(\alpha)v_{y},
\\
\frac{d(v_{y}/v_{x})}{d\ell} &=& \left[f_{B}(\alpha) -
f_{A}(\alpha)\right]\frac{v_{y}}{v_{x}},
\\
\frac{d(tv_{x})}{d \ell}&=&0,
\\
\frac{d t}{d\ell}&=&-f_{A}(\alpha)t,
\\
\frac{d\alpha}{d\ell}&=&-\frac{1}{2}f_{C}(\alpha)\alpha.
\end{eqnarray}
Notice that $\lambda=\lambda_{0}=0$ has been taken. As shown in
Fig.~\ref{Fig:VRGClean}, both $v_{x}$ and $v_{y}$ increase slowly
with growing $\ell$, and $\alpha$ flows to zero slowly in the lowest
energy limit, implying that Coulomb interaction is marginally
irrelevant.

Parameter $t$ vanishes as $\ell \rightarrow \infty$, thus the
Coulomb interaction suppresses the tilt, consistent with previous
works \cite{Isobe12, LeeLee18}. Detassis \emph{et al.}
\cite{Detassis17} argued that this conclusion is also applicable to
3D tilted WSM. As the effective tilt goes to zero, the fermion
velocities display asymptotically the same behavior as the untilted
case: the velocities increase logarithmically as the energy scale is
lowering. The results are shown in Figs.~\ref{Fig:VRGClean}~(a) and
(b). For free 2D Dirac fermions, the DOS, specific heat, and
compressibility behave as $\rho(\omega)\sim\omega/(v_{x}v_{y})$,
$C_{v}(T)\sim T^{2}/(v_{x}v_{y})$, and $\kappa(T)\sim
T/(v_{x}v_{y})$, respectively. Once the singular renormalization of
fermion velocities are incorporated, the DOS, specific heat, and
compressibility of interacting Dirac fermions \cite{Kotov12,
Vafek07, Sheehy07} become $\rho(\omega) \sim
\omega/\ln^{2}(\omega_{0}/\omega)$, $C_{v}(T) \sim
T^{2}/\ln^{2}\left(T_{0}/T\right)$, and $\kappa(T) \sim
T/\ln^{2}(T_{0}/T)$, respectively. These results are already known
previously.

\subsection{Disorder effects \label{SubSec:NumResutlsDisorder}}

We next investigate the physical effect of disorders on the behavior
of non-interacting tilted Dirac fermions. Under RG transformations,
the disorder may be irrelevant, marginally irrelevant, marginal, and
relevant as the running scale $\ell$ increases. For an irrelevant
disorder, the effective strength flows to zero rapidly, and the
low-energy properties of the system is not qualitatively changed by
disorder scattering. For a marginally irrelevant disorder, the
effective strength vanishes slowly, and the observable quantities
acquire weak logarithmic-like corrections to their energy or
temperature dependence. For a marginal disorder, the effective
strength flows to a fixed point, and the observable quantities
receives power-law corrections. For a relevant disorder, the
effective strength increases indefinitely with growing $\ell$, and
thus the system becomes unstable and should enter into a distinct
phase. As demonstrated in extensive RG studies, relevant disorder
can convert the SM into a CDM phase \cite{Ludwig94, Ostrovsky06,
Evers08, Foster12, Goswami11, Nandkishore17, YuXuanWang17,
WangLiuZhang17MWSM, Mandal17, Sikkenk17, Roy14, Syzranov16, Roy16,
Luo18, Roy18}, in which the fermions acquire a finite disorder
scattering rate $\gamma_{0}$. In addition, the zero-energy fermion
DOS $\rho(0)$ takes a finite value that is determined by
$\gamma_{0}$. In contrast, $\rho(0)=0$ in the SM phase. Thus, the SM
and CDM phases can be well distinguished by the value of
$\gamma_{0}$ and $\rho(0)$. In this subsection, we present the
detailed RG results for RSP, $x$-RVP, $y$-RVP, and RM, and analyze
the unusual disorder-induced properties.

\begin{figure}[htbp]
\center
\includegraphics[width=3.35in]{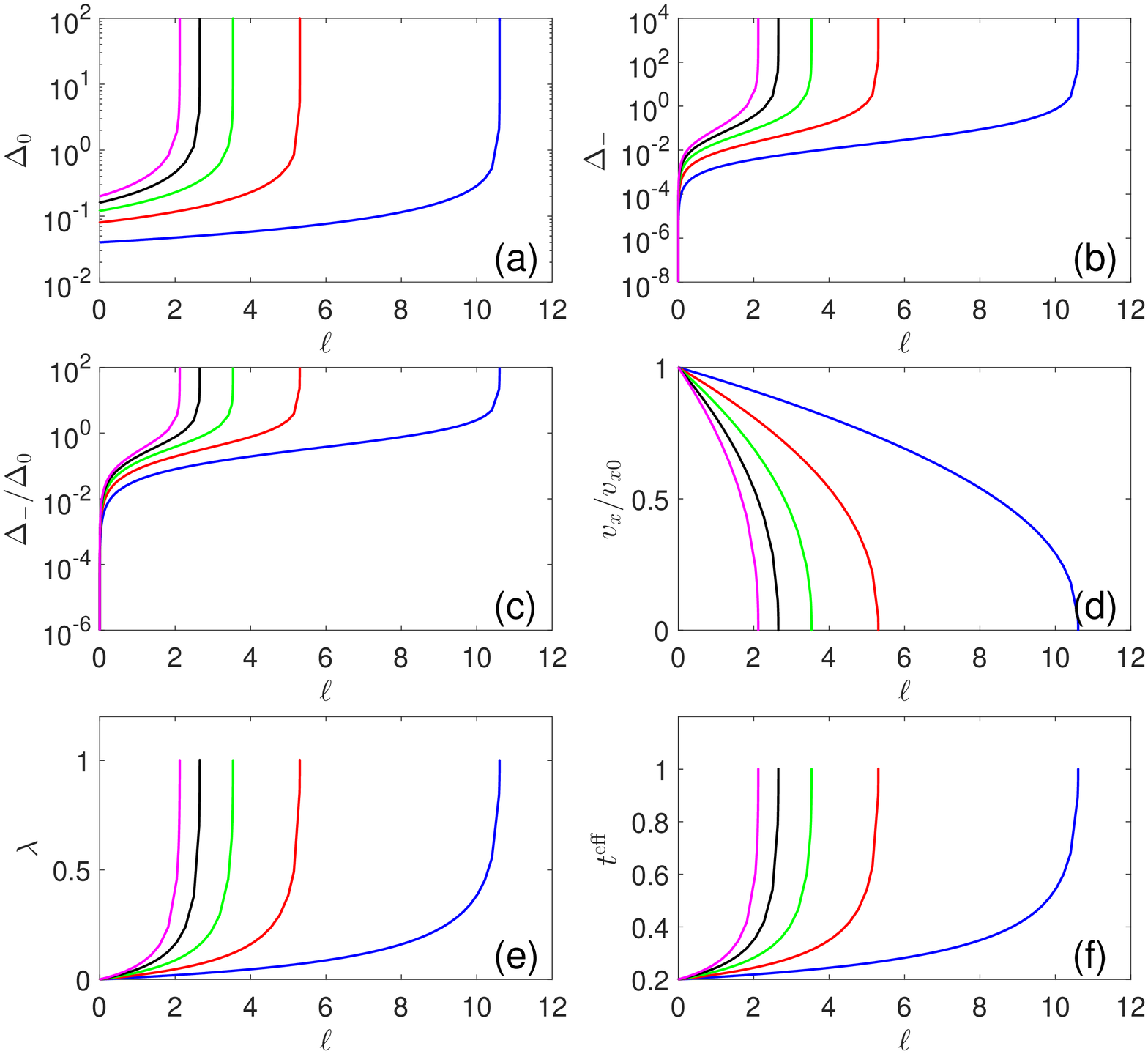}
\caption{RG flows of $\Delta_{0}$, $\Delta_{-}$,
$\Delta_{-}/\Delta_{0}$, and $t^{\mathrm{eff}}$ due to RSP. Blue,
red, green, black, and magenta lines correspond to initial values
$\Delta_{0,0}=0.04, 0.08, 0.12, 0.16, 0.2$, respectively. Here,
$t_{0}=0.2$ and $v_{y0}/v_{x0}=1$. \label{Fig:VRGIniRSP}}
\end{figure}

\subsubsection{RSP}

We first consider RSP. Its strength parameter $\Delta_{0}$ flows to
infinity at a finite scale $\ell_c$, as showed by
Fig.~\ref{Fig:VRGIniRSP}(a). A unique feature of RSP is that it
always generates the fermion-disorder coupling term defined by
Eq.~(\ref{Eq:SdisMinus}) if the tilt parameter $t_0$ is nonzero.
According to Fig.~\ref{Fig:VRGIniRSP}(b), the strength parameter
$\Delta_{-}$ also diverges rapidly as $\ell \rightarrow \ell_c$.
From Fig.~\ref{Fig:VRGIniRSP}(c), we observe that the ratio
$\Delta_{-}/\Delta_{0}\rightarrow \infty$ as $\ell$ increases. This
indicates that the dynamically generated disorder is much more
important than RSP at low energies. The divergence of disorder
strength parameters indicates that the tilted DSM becomes unstable
and would be turned into a CDM in which both $\gamma_0$ and
$\rho(0)$ are nonzero. The tilt parameter $t$ is fixed at its
initial value, namely $t = t_{0}$. Interestingly, we find that
$\lambda \rightarrow 1$ as $\ell$ grows as shown in
Fig.~\ref{Fig:VRGIniRSP}(e). Consequently, the effective tilt
parameter now becomes
\begin{eqnarray}
t^{\mathrm{eff}} = \frac{t+\lambda}{1+t\lambda} \rightarrow
\frac{t_{0}+1}{1+t_{0}} = 1,
\end{eqnarray}
which is displayed in Fig.~\ref{Fig:VRGIniRSP}(f).

The eigenvalues of a Hamiltonian contain important information. For
an interacting fermion system, the real part of the energy
eigenvalues represents the energy dispersion of fermions, whereas
the imaginary part characterizes the fermion damping effect. For a
non-interacting 2D DSM, the real part of the energy vanishes at
discrete points in the Brillouin zone, corresponding to the Dirac
points, and the imaginary part is zero. Under certain circumstances,
the real part of the energy may vanish along a finite curve in the
Brillouin zone \cite{Kozii17, Papaj18} after incorporating the
corrections due to disorder scattering, electron-electron scatting,
or electron-phonon scattering. This curve is usually called a bulk
Fermi arc, which has recently attracted considerable research
interest \cite{Kozii17, Papaj18, Zhao18}.

We now examine whether RSP leads to a bulk Fermi arc in the system
under consideration. Once finite disorder scattering rate
$\gamma_{0}$ is generated in the CDM phase, the retarded fermion
propagator can be written as
\begin{widetext}
\begin{eqnarray}
G^{R}(\omega,\mathbf{k}) = \frac{1}{\omega+i\gamma_{0} -
tv_{x}k_{x}- (\lambda\omega+it\gamma_{0}) \sigma_{x} -
v_{x}k_{x}\sigma_{x}-v_{y}k_{y}\sigma_{y}}.
\end{eqnarray}
The eigenvalues of disordered Hamiltonian are determined by
\begin{eqnarray}
&&\left|\left|\left(E+i\gamma_{0}\right) - tv_{x}k_{x}-\left(
\lambda E + it\gamma_{0}\right)\sigma_{x} -
v_{x}k_{x}\sigma_{x}-v_{y}k_{y}\sigma_{y}\right|\right|=0,
\end{eqnarray}
which is equivalent to
\begin{eqnarray}
&&\left|
\begin{array}{cc}
E+i\gamma_{0}-tv_{x}k_{x} & - \left(\lambda
E+it\gamma_{0}\right)-v_{x}k_{x}-iv_{y}k_{y}
\\
-\left(\lambda E+it\gamma_{0}\right) - v_{x}k_{x}+iv_{y}k_{y} &
E+i\gamma_{0}-tv_{x}k_{x}
\end{array}
\right|=0.
\end{eqnarray}
The solution of this equation is
\begin{eqnarray}
E_{\pm}(\mathbf{k}) = \frac{\left(t+\lambda\right) v_{x}k_{x} -
\left(1-t\lambda\right)i\gamma_{0}}{1-\lambda^{2}}
\pm\sqrt{\frac{1}{\left(1-\lambda^{2}\right)^{2}}\left[\left(1 +
t\lambda\right)v_{x}k_{x}+\left(t-\lambda\right)i\gamma_{0}\right]^{2}
+ \frac{1}{1-\lambda^{2}}v_{y}^{2}k_{y}^{2}}.
\label{Eq:DispersionCDMA}
\end{eqnarray}
\end{widetext}
If $\lambda$ flows to a fixed point $\lambda = t$, the scattering
rate $\gamma_{0}$ does not induce bulk Fermi arc, but only
represents fermion damping. However, our RG analysis shows that
generically $\lambda\ne t$. Thus, for $k_{x}=0$,
\begin{eqnarray}
E_{\pm}(\mathbf{k}) = \frac{\left(t\lambda-1\right)
i\gamma_{0}}{1-\lambda^{2}} \pm\sqrt{\frac{v_{y}^{2}
k_{y}^{2}}{1-\lambda^{2}}-\gamma_{0}^{2} \left(\frac{t -
\lambda}{1-\lambda^{2}}\right)^2}. \label{Eq:DispersionCDMB}
\end{eqnarray}
If $|k_{y}| < \frac{|t-\lambda|}{v_{y} \sqrt{1-\lambda^{2}}}
\gamma_{0}$, the energy $E$ is pure imaginary. There emerges a bulk
Fermi arc in the Brillouin zone \cite{Kozii17, Papaj18}, which
replaces Dirac points. In this state, the fermion DOS, specific
heat, and compressibility behave as $\rho(0)>0$, $C_{v}(T)\sim T$,
and $\kappa(0)>0$. If the Dirac cone is not tilted, i.e., $t=0$, we
always have $\lambda = \lambda_{0}=0$. Although RSP still leads to
CDM transition in the untilted case, there is no bulk Fermi arc.

\subsubsection{$x$-RVP}

\begin{figure}[htbp]
\center
\includegraphics[width=3.35in]{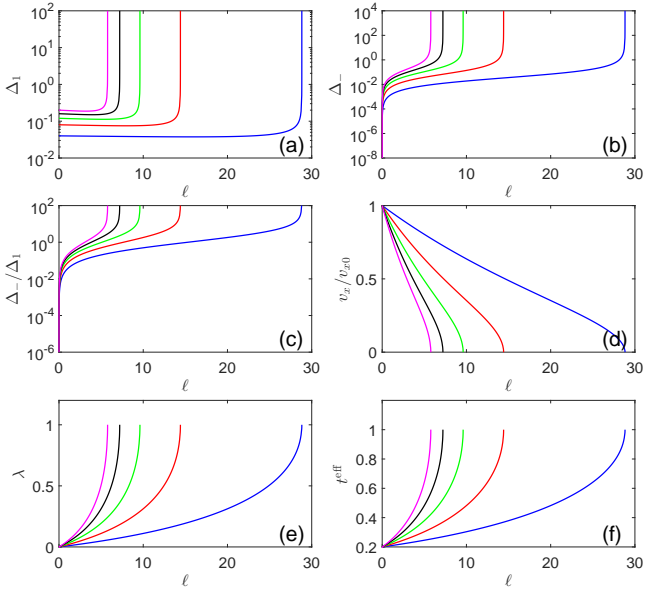}
\caption{RG flows of $\Delta_{1}$, $\Delta_{-}$,
$\Delta_{-}/\Delta_{1}$, and $t^{\mathrm{eff}}$ due to $x$-RVP.
Blue, red, green, black, and magenta lines correspond to the initial
values $\Delta_{1,0}=0.04, 0.08, 0.12, 0.16, 0.2$, respectively.
Here, $t_{0}=0.2$ and $v_{y0}/v_{x0}=1$.} \label{Fig:VRGIniRVPx}
\end{figure}

In case $x$-RVP exists by itself, we plot the flows of $\Delta_{1}$,
$\Delta_{-}$, $\Delta_{-}/\Delta_{1}$, $v_{x}$, $\lambda$, and $t$
in Figs.~\ref{Fig:VRGIniRVPx}(a)-(f). The results are qualitatively
the same as those shown in Fig.~\ref{Fig:VRGIniRSP}. The $x$-RVP
also generates the new fermion-disorder coupling term
Eq.~(\ref{Eq:DisorderGenerateTerm}), which then leads to a finite
scattering rate $\gamma_{0}$ as well as a bulk Fermi arc.

We emphasize that the emergence of bulk Fermi arc induced by RSP or
$x$-RVP is closely related to the presence of the term
Eq.~(\ref{Eq:DisorderGenerateTerm}). If this term is naively
discarded in the calculation, we would find that $\lambda$ always
flows to the fixed point $\lambda = t$. In this case, no bulk Fermi
arc emerges even though a finite $\gamma_0$ is generated. The
detailed analysis is presented in
Appendix~\ref{App:RGResultsDisgardingImportantTerm}. Here, we
briefly discuss the RG flow of $\lambda$. We already see from
Figs.~\ref{Fig:VRGIniRSP} and \ref{Fig:VRGIniRVPx} that
$\lambda\rightarrow 1$ and $t^{\mathrm{eff}}\rightarrow1$ at certain
scale $\ell_{c}$ due to RSP or $x$-RVP. The flow equation for
$\lambda$ is
\begin{eqnarray}
\frac{d\lambda}{dt} &=& \frac{1+t\lambda}{\left(1 -
t^{2}\right)^{3/2}}\left[\left(t-\lambda\right)
\left(\Delta_{0}+\Delta_{1}\right) \right.\nonumber
\\
&&\left.+ (1-t)\left(1-\lambda\right)\frac{\Delta_{-}}{2}\right].
\label{Eq:VRGLambdaBoth}
\end{eqnarray}
Dropping disorder $\Delta_{-}$, this flow equation becomes
\begin{eqnarray}
\frac{d\lambda}{dt} = \frac{1+t\lambda}{\left(1-t^{2}\right)^{3/2}}
\left(t-\lambda\right)\left(\Delta_{0}+\Delta_{1}\right).
\label{Eq:VRGLambdaOld}
\end{eqnarray}
It is easy to see that $\lambda=t$ is a fixed point. thus $\lambda
\rightarrow t$ if we ignore $\Delta_{-}$. If we choose to drop
$\Delta_{0}$ and $\Delta_{1}$, the flow equation would become
\begin{eqnarray}
\frac{d\lambda}{dt} = \frac{1+t\lambda}{\left(1-t^{2}\right)^{3/2}}
\left(1-t\right)\left(1-\lambda\right)\frac{\Delta_{-}}{2}.
\label{Eq:VRGLambdaNew}
\end{eqnarray}
In this limit, $\lambda = 1$ is a fixed point, and $\lambda
\rightarrow 1$ as $\ell$ increases. Since
$\Delta_{-}/\Delta_{0}\rightarrow \infty$ and
$\Delta_{-}/\Delta_{1}\rightarrow \infty$, the dynamically generated
disorder dominates over RSP and $x$-RVP in the low-energy region.
Comparing to $\Delta_{-}$, $\Delta_{0}$ and $\Delta_{1}$ could be
asymptotically neglected. Therefore, one can approximately replace
Eq.~(\ref{Eq:VRGLambdaBoth}) with Eq.~(\ref{Eq:VRGLambdaNew}). This
is the reason why $\lambda\rightarrow 1$. Sikkenk and Fritz
\cite{Sikkenk17} studied the disorder effects in tilted 3D WSM, and
found that the parameter $\lambda \rightarrow 1$ as the disorder
strength parameter flows to infinity. This is well consistent with
our results obtained in tilted 2D DSM with RSP or $x$-RVP.

\subsubsection{$y$-RVP}

The $x$- and $y$-components of RVP are equivalent in untilted
systems. But they become distinct if the Dirac cone is tilted along
$x$-axis. When $y$-RVP is added to tilted 2D DSM, it does not
generate new disorder, and the corresponding RG equations are
\begin{eqnarray}
\frac{dv_{x}}{d\ell}&=& -\frac{1+t\lambda}{\left(1 -
t^{2}\right)^{3/2}}\Delta_{2}v_{x},
\\
\frac{dv_{y}}{d\ell} &=& -\frac{1+t\lambda}{\left(1 -
t^{2}\right)^{3/2}}\Delta_{2}v_{y},
\\
\frac{d(v_{y}/v_{x})}{d\ell}&=&0,
\\
\frac{d(tv_{x})}{d \ell} &=& -\frac{1+t\lambda}{\left(1 -
t^{2}\right)^{3/2}}\Delta_{2}tv_{x},
\\
\frac{d t}{d\ell}&=&0,
\\
\frac{d\lambda}{d\ell} &=& -\frac{\left(1+t\lambda\right)
\left(\lambda+t\right)}{\left(1-t^{2}\right)^{3/2}}\Delta_{2},
\\
\frac{d\Delta_{2}}{d\ell}&=& 0.
\end{eqnarray}

\begin{figure}[htbp]
\center
\includegraphics[width=3.35in]{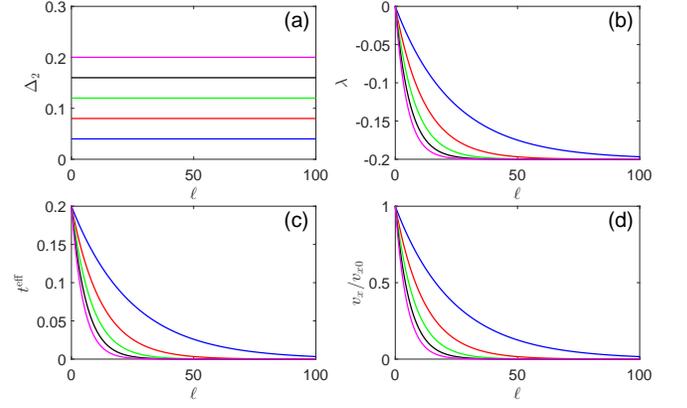}
\caption{Flows of $\Delta_{2}$, $\lambda$, $t^{\mathrm{eff}}$, and
$v_{x}$ due to $y$-RVP. Blue, red, green, black, and magenta lines
correspond to the initial values $\Delta_{2,0}=0.04, 0.08, 0.12,
0.16, 0.2$, respectively. Here, $t_{0}=0.2$ and $v_{y0}/v_{x0}=1$.}
\label{Fig:VRGIniRVPy}
\end{figure}

The tilt $t$, velocity ratio $v_{y}/v_{x}$, and disorder strength
$\Delta_{2}$ are all independent of $\ell$, and thus can be fixed at
constants, namely $t = t_{0}$, $v_{y}/v_{x} = v_{y0}/v_{x0}$, and
$\Delta_{2} = \Delta_{2,0}$. Then the RG equation for $\lambda$
becomes
\begin{eqnarray}
\frac{d\lambda}{d\ell}&=&-\frac{\left(1+t_{0}\lambda\right)
\left(\lambda+t_{0}\right)}{\left(1-t_{0}^{2}\right)^{3/2}}\Delta_{2,0}.
\end{eqnarray}
For initial value $\lambda_{0} = 0$, $\lambda$ flows quickly to a
stable fixed point $\lambda^{*} = -t_{0}$. In the lowest energy
limit, the effective tilt parameter satisfies
\begin{eqnarray}
{t^{\mathrm{eff}}}^{*} = \frac{t_{0} + \lambda^{*}}{1 +
t_{0}\lambda^{*}} = 0.
\end{eqnarray}
Thus $y$-RVP tends to suppress the tilt. The RG flows of
$\Delta_{2}$, $\lambda$, $t^{\mathrm{eff}}$, and $v_{x}$ with
varying $\ell$ are shown in Fig.~\ref{Fig:VRGIniRVPy}.

In the low-energy region, the RG equations for $v_{x}$ and $v_{y}$
are approximately given by
\begin{eqnarray}
\frac{d v_{x}}{d \ell} &\sim& - \frac{1+t_{0}\lambda^{*}}{\left(1 -
t_{0}^{2}\right)^{3/2}}\Delta_{2,0}v_{x}\sim-\eta_{0}v_{x},
\\
\frac{d v_{y}}{d \ell} &\sim& -\frac{1+t_{0}\lambda^{*}}{\left(1 -
t_{0}^{2}\right)^{3/2}}\Delta_{2,0}v_{y}\sim-\eta_{0}v_{y},
\end{eqnarray}
where
\begin{eqnarray}
\eta_{0} = \frac{\Delta_{2,0}}{\left(1-t_{0}^{2}\right)^{1/2}}.
\label{Eq:eta0Def}
\end{eqnarray}
The solutions of $v_{x}$ and $v_{y}$ are
\begin{eqnarray}
v_{x}\sim v_{x0}e^{-\eta_{0}\ell},\qquad
v_{y}\sim v_{y0}e^{-\eta_{0}\ell}.
\end{eqnarray}
It is clear that $v_{x}$ and $v_{y}$ approach to zero as
$\ell\rightarrow \infty$. Employing the transformation $k =
k_{0}e^{-\ell}$, where $k_{0}$ is taken as a fixed large value of
$k$, we further express $v_{x}$ and $v_{y}$ as
\begin{eqnarray}
v_{x}\sim v_{x0}\left(\frac{k}{k_{0}}\right)^{\eta_{0}},\qquad
v_{y}\sim v_{y0}\left(\frac{k}{k_{0}}\right)^{\eta_{0}}.
\end{eqnarray}
The parameters $v_{x}^{\mathrm{eff}}$ and $v_{y}^{\mathrm{eff}}$ are
given by
\begin{eqnarray}
v_{x}^{\mathrm{eff}} &\sim& \frac{1+t_{0}
\lambda^{*}}{1-{\lambda^{*}}^{2}}v_{x} \sim
v_{x0}
\left(\frac{k}{k_{0}}\right)^{\eta_{0}}, \\
v_{y}^{\mathrm{eff}}&\sim&\frac{1}{\sqrt{1-{\lambda^{*}}^{2}}}
v_{y}\sim \frac{1}{\sqrt{1-t_{0}^{2}}}
v_{y0}\left(\frac{k}{k_{0}}\right)^{\eta_{0}}.
\end{eqnarray}

\begin{figure}[htbp]
\center
\includegraphics[width=3.35in]{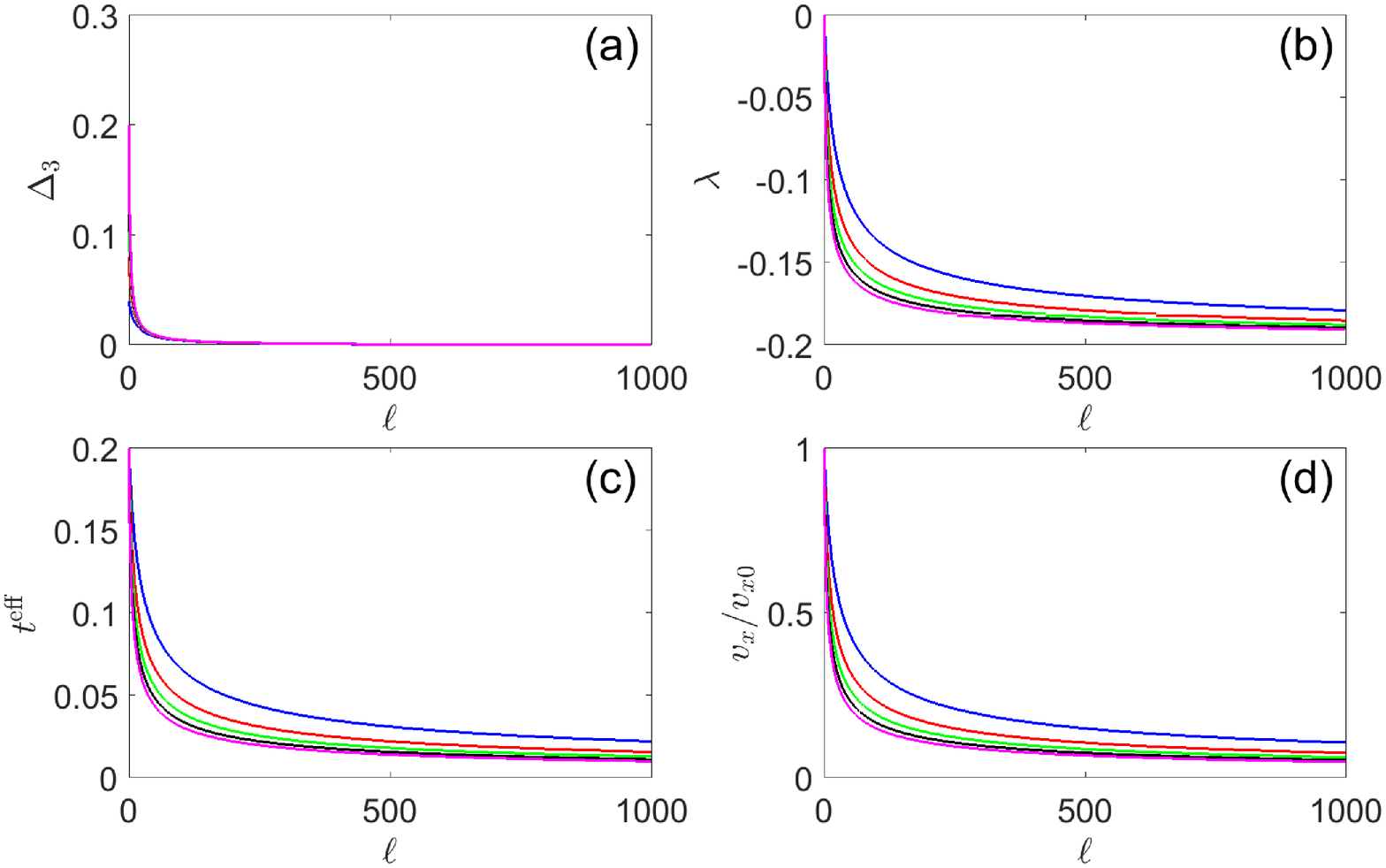}
\caption{Flows of $\Delta_{3}$, $\lambda$, $t^{\mathrm{eff}}$, and
$v_{x}$ due to RM. Blue, red, green, black, and magenta lines
correspond to the initial values $\Delta_{3,0}=0.04, 0.08, 0.12,
0.16, 0.2$, respectively. Here, $t_{0}=0.2$ and $v_{y0}/v_{x0}=1$.}
\label{Fig:VRGIniRM}
\end{figure}

We notice that the dynamical exponent $z$ becomes $z=1+\eta_{0}$.
Accordingly, the DOS depends on $\omega$ as follows
\begin{eqnarray}
\rho(\omega)\sim\omega^{\frac{d}{z}-1}\sim \omega^{\frac{2}{1 +
\eta_{0}}-1}\sim\omega^{\frac{1-\eta_{0}}{1+\eta_{0}}}.
\end{eqnarray}
The specific heat and compressibility depend on $T$ as
\begin{eqnarray}
C_{v}(T)\sim T^{\frac{d}{z}}\sim T^{\frac{2}{1+\eta_{0}}},
\\
\kappa(T)\sim T^{\frac{d}{z}-1}\sim
T^{\frac{1-\eta_{0}}{1+\eta_{0}}}.
\end{eqnarray}
These three quantities acquire power-law corrections. Comparing to
the clean case, they are all enhanced by $y$-RVP. For a given
$\Delta_{2,0}$, $\eta_{0}$ becomes larger with growing of $t_{0}$,
as shown in Eq.~(\ref{Eq:eta0Def}), and the enhancement of DOS,
specific heat, and compressibility is more significant. This
indicates the influence of $y$-RVP is amplified by the tilt of Dirac
cone.

\subsubsection{RM}

Similar to $y$-RVP, RM also does not generate new disorders. The
corresponding RG equations are
\begin{eqnarray}
\frac{d v_{x}}{d\ell} &=& -\frac{1+t\lambda}{\left(1-t^{2}
\right)^{3/2}}\Delta_{3}v_{x}, \\
\frac{d v_{y}}{d\ell} &=& -\frac{1+t\lambda}{\left(1 -
t^{2}\right)^{3/2}}\Delta_{3}v_{y}, \\
\frac{d(v_{y}/v_{x})}{d\ell} &=& 0, \label{Eq:VRGVRatioRM}
\\
\frac{d(tv_{x})}{d \ell} &=& -\frac{1+t\lambda}{\left(1-t^{2}
\right)^{3/2}}\Delta_{3}t v_{x},
\\
\frac{d t}{d\ell}&=&0, \label{Eq:VRGwRM}
\\
\frac{d\lambda}{d\ell}&=&-\frac{\left(1+t\lambda\right)
\left(\lambda+t\right)}{\left(1-t^{2}\right)^{3/2}}\Delta_{3},
\\
\frac{d\Delta_{3}}{d\ell}&=&-2\frac{1+t\lambda}{\left(1 -
t^{2}\right)^{3/2}}\Delta_{3}^{2}.
\end{eqnarray}
According to Eq.~(\ref{Eq:VRGVRatioRM}) and Eq.~(\ref{Eq:VRGwRM}),
we set $t = t_{0}$ and $v_{y}/v_{x} = v_{y0}/v_{x0}$. We solve the
rest flow equations at initial value $\lambda = 0$, and display the
numerical results in Fig.~\ref{Fig:VRGIniRM}. In the lowest energy
limit, we find that
\begin{eqnarray}
\lambda&\rightarrow&\lambda^{*} = -t_{0},
\\
\Delta_{3} &\rightarrow& 0, \\
{t^{\mathrm{eff}}}^{*} &=& \frac{t_{0} + \lambda^{*}}{1 +
t_{0}\lambda^{*}} = 0.
\end{eqnarray}
Therefore, RM forces the tilted Dirac cone to go back to the
untilted limit, in close analogy to the case of $y$-RVP. The RG
equation for $\Delta_{3}$ is re-written as
\begin{eqnarray}
\frac{d\Delta_{3}}{d\ell} &\sim& -2\frac{1}{\left(1 -
t_{0}^{2}\right)^{1/2}}\Delta_{3}^{2}.
\end{eqnarray}
Its solution is
\begin{eqnarray}
\Delta_{3}\sim\frac{\Delta_{3,0}}{1+2\left(1-t_{0}^{2}\right)^{-1/2}
\Delta_{3,0}\ell}, \label{Eq:Delta3SolutionOnlyRM}
\end{eqnarray}
which approaches to zero slowly as $\ell\rightarrow \infty$.

In the low-energy region, one can approximate the RG equations for
$v_{x}$ and $v_{y}$ by
\begin{eqnarray}
\frac{d v_{x}}{d \ell}&\sim& -\frac{1+t_{0}\lambda^{*}}{\left(1 -
t_{0}^{2}\right)^{3/2}}\Delta_{3}v_{x} \sim
-\frac{\Delta_{3}}{\left(1-t_{0}^{2}\right)^{1/2}}v_{x},
\label{Eq:RGEqVxApproOnlyRM}
\\
\frac{d v_{y}}{d \ell}&\sim& - \frac{1+t_{0}\lambda^{*}}{\left(1 -
t_{0}^{2}\right)^{3/2}}\Delta_{3}v_{y} \sim -
\frac{\Delta_{3}}{\left(1-t_{0}^{2}\right)^{1/2}}v_{y}.
\label{Eq:RGEqVyApproOnlyRM}
\end{eqnarray}
Substituting Eq.~(\ref{Eq:Delta3SolutionOnlyRM}) into
Eqs.~(\ref{Eq:RGEqVxApproOnlyRM}) and (\ref{Eq:RGEqVyApproOnlyRM}),
we find that $v_{x}$ and $v_{y}$ behave as
\begin{eqnarray}
v_{x} &\sim& \frac{v_{x0}}{\sqrt{1 + 2\left(1 -
t_{0}^{2}\right)^{-3/2}\Delta_{3,0}\ell}}, \\
v_{y} &\sim& \frac{v_{y0}}{\sqrt{1+2\left(1 -
t_{0}^{2}\right)^{-3/2}\Delta_{3,0}\ell}}.
\end{eqnarray}
Both $v_{x}$ and $v_{y}$ flow to zero slowly with growing $\ell$.
Making use of the transformation $k=k_{0}e^{-\ell}$, we obtain
\begin{eqnarray}
\frac{v_{x}}{v_{x0}}\sim\frac{v_{y}}{v_{y0}} \sim
\frac{1}{\sqrt{1+2\left(1-t_{0}^{2}\right)^{-3/2}
\Delta_{3,0}\ln\left(\frac{k_{0}}{k}\right)}}.
\end{eqnarray}

\begin{figure*}[htbp]
\center
\includegraphics[width=6.9in]{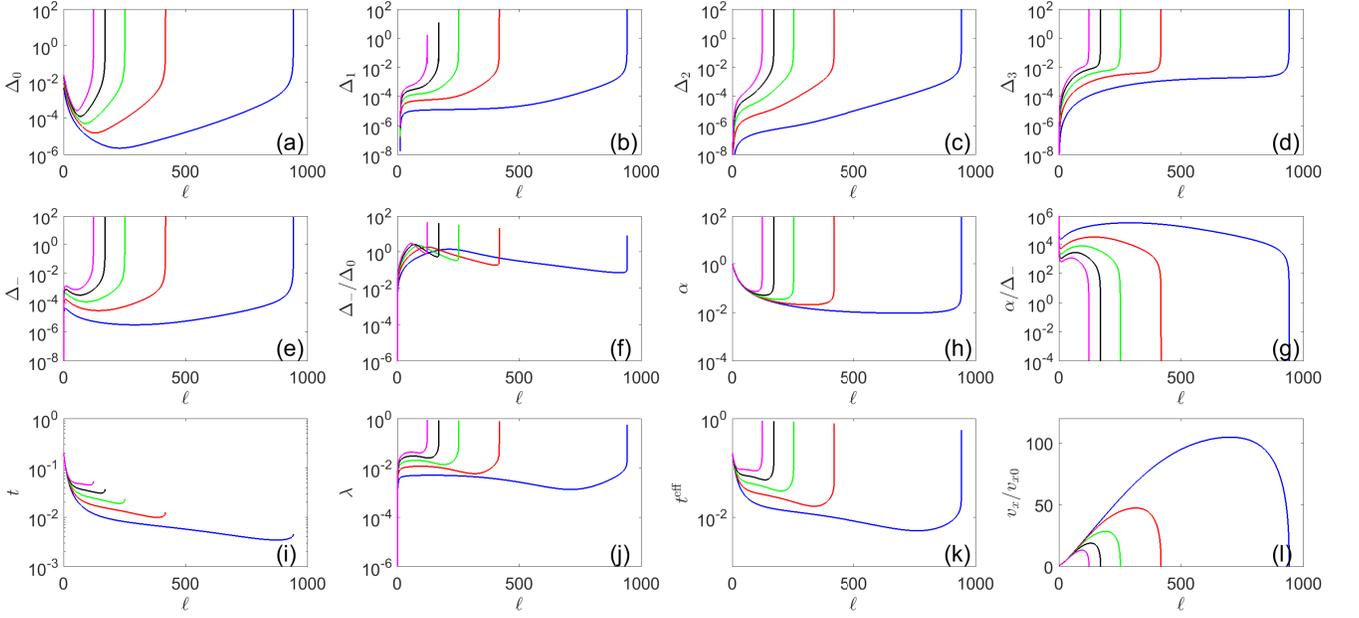}
\caption{Flows of $\Delta_{0}$, $\Delta_{1}$, $\Delta_{2}$,
$\Delta_{3}$, $\Delta_{-}$, $\Delta_{-}/\Delta_{0}$, $\alpha$,
$\alpha/\Delta_{0}$, $t$, $\lambda$, $t^{\mathrm{eff}}$, and $v_{x}$
due to the interplay of RSP and Coulomb interaction. Blue, red,
green, black, and magenta lines correspond to initial values
$\Delta_{0,0}=0.005, 0.01, 0.015, 0.02, 0.025$, respectively. Here,
$t_{0}=0.2$ and $v_{y0}/v_{x0}=1$. The flavor is $N = 4$.}
\label{Fig:VRGIniRSPLRC}
\end{figure*}

The parameters $v_{x}^{\mathrm{eff}}$ and $v_{y}^{\mathrm{eff}}$ are
approximated as
\begin{eqnarray}
v_{x}^{\mathrm{eff}} &\sim& \frac{1+t_{0} \lambda^{*}}{1 -
{\lambda^{*}}^{2}}v_{x} \sim v_{x},
\\
v_{y}^{\mathrm{eff}}&\sim&\frac{1}{\sqrt{1-{\lambda^{*}}^{2}}}
v_{y}\sim \frac{1}{\sqrt{1-t_{0}^{2}}}v_{y}.
\end{eqnarray}
We see that $v_{x}^{\mathrm{eff}}$ and $v_{y}^{\mathrm{eff}}$
exhibit the same momentum dependence as $v_{x}$ and $v_{y}$,
respectively. In the clean limit, the DOS, specific heat, and
compressibility depend on $\omega$ or $T$ as: $\rho(\omega)\sim
\omega/(v_{x}^{\mathrm{eff}} v_{y}^{\mathrm{eff}})$,
$C_{v}(T)\sim T^{2}/(v_{x}^{\mathrm{eff}}
v_{y}^{\mathrm{eff}})$, and $\kappa(T)\sim
T/(v_{x}^{\mathrm{eff}}v_{y}^{\mathrm{eff}})$. After
considering the RM-induced corrections, these three quantities
become
\begin{eqnarray}
\rho(\omega)&\sim& \omega\ln\left(\frac{\omega_{0}}{\omega}\right),
\\
C_{v}(T)&\sim& T^{2}\ln\left(\frac{T_{0}}{T}\right),
\\
\kappa(T)&\sim& T\ln\left(\frac{T_{0}}{T}\right),
\end{eqnarray}
which display logarithmic corrections. Therefore, although RM and
$y$-RVP both suppress tilt, they result in distinct low-energy
properties of Dirac fermions.

\subsection{Interplay between interaction and disorder \label{SubSec:NumResutlsInterplay}}

The results of Sec.~\ref{SubSec:NumResutlsDisorder} are obtained in
the non-interacting limit. We now turn to study the interplay
between the Coulomb interaction and each single type of disorder,
with the purpose of determining the physical consequence of Dirac
cone tilt in realistic 2D DSMs.

\subsubsection{Coulomb interaction and RSP}

When the Coulomb interaction and RSP are both present, they
automatically generate all the other types of disorder, including
$x$-RVP, $y$-RVP, RM, and the new disorder described by
Eq.~(\ref{Eq:SdisMinus}). As shown in Fig.~\ref{Fig:VRGIniRSPLRC},
all the disorder strength parameters diverge at a finite scale
$\ell_c$. The Coulomb interaction strength parameter $\alpha$ is
also divergent at this scale, but the ratio $\alpha/\Delta_{i}$
vanishes. An apparent fact is that disorder always dominates over
the Coulomb interaction at low energies, and determines the
low-energy behaviors of the system. Consequently, there is always a
finite disorder scattering rate and a bulk Fermi arc in the
Brillouin zone.

The combination of Coulomb interaction and RSP has already been
studied in the context of untilted 2D DSM \cite{Stauber05,
WangLiu14}. While RSP is more important than weak Coulomb
interaction and triggers the SM-to-CDM phase transition, a
sufficiently strong Coulomb interaction can substantially suppress
RSP and restore the original SM state. However, the SM state cannot
be restored by the strong Coulomb interaction in tilted 2D DSM. It
is therefore clear that the tilt does give rise to different
properties than the untilted case.

\subsubsection{Coulomb interaction and $x$-RVP}

Similar to RSP, the coexistence of Coulomb interaction and $x$-RVP
also generates all the other types of disorder. The interaction
strength $\alpha$ and the disorder parameters $\Delta_{i}$ also flow
to the strong coupling regime, and their ratio $\alpha/\Delta_{i}$
still goes to zero. Thus, the system is inevitably turned into a CDM
phase that features a finite scattering rate $\gamma_0$.
Additionally, there also emerges a bulk Fermi arc. The model
parameters depend on $\ell$ in qualitatively the same way as
Fig.~\ref{Fig:VRGIniRSPLRC}, and thus are not shown.

\begin{figure}[htbp]
\center
\includegraphics[width=3.35in]{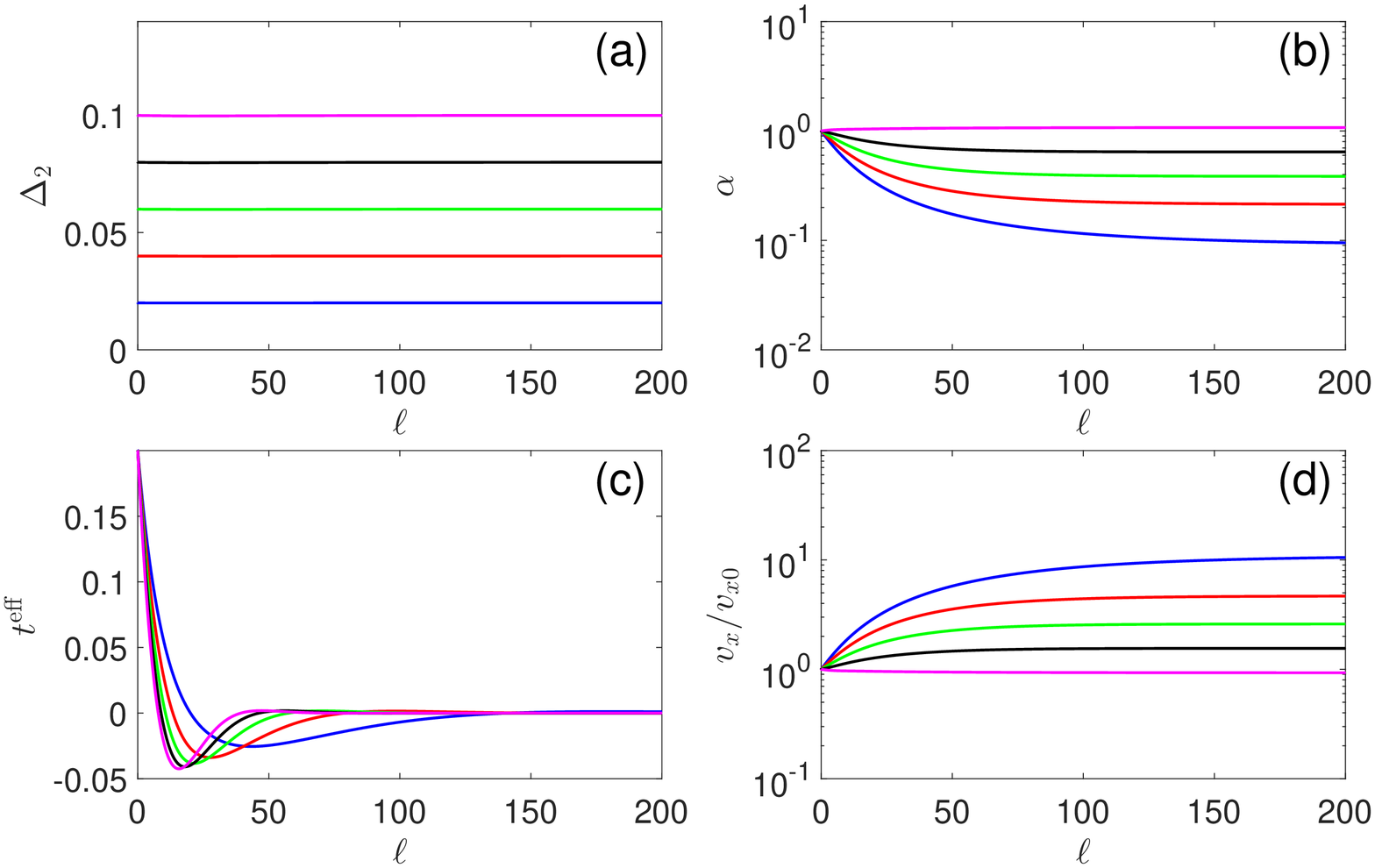}
\caption{Flows of $\Delta_{2}$, $\lambda$, $t^{\mathrm{eff}}$, and
$v_{x}$ due to the interplay between $y$-RVP and Coulomb
interaction. Blue, red, green, black, and magenta lines correspond
to initial values $\Delta_{2, 0}=0.02, 0.04, 0.06, 0.08, 0.1$,
respectively. Here, $t_{0}=0.2$, $v_{y0}/v_{x0}=1$, and $N=4$.}
\label{Fig:VRGIniRVPyLRC}
\end{figure}

\subsubsection{Coulomb interaction and $y$-RVP}

Coulomb interaction and $y$-RVP combine to yield
\begin{eqnarray}
\frac{d v_{x}}{d\ell}&=& \left[\mathcal{H}_{1}(\alpha) -
\frac{1+t\lambda}{\left(1-t^{2}\right)^{3/2}}\Delta_{2}\right]v_{x},
\\
\frac{d v_{y}}{d \ell}&=& \left[\mathcal{H}_{2}(\alpha) -
\frac{1+t\lambda}{\left(1-t^{2}\right)^{3/2}}\Delta_{2}\right]v_{y},
\\
\frac{d(v_{y}/v_{x})}{d\ell} &=& \left[\mathcal{H}_{2}(\alpha) -
\mathcal{H}_{1}(\alpha)\right] \frac{v_{y}}{v_{x}},
\\
\frac{d(tv_{x})}{d\ell} &=& \left[-\frac{1+t\lambda}{\left(1 -
t^{2}\right)^{3/2}}\Delta_{2}t+\mathcal{H}_{3}(\alpha)\right]v_{x},
\\
\frac{d t}{d\ell} &=& \mathcal{H}_{3}(\alpha) -
\mathcal{H}_{1}(\alpha)t,
\\
\frac{d\lambda}{d\ell} &=& -\frac{\left(1+t\lambda\right)
\left(\lambda + t\right)}{\left(1-t^{2}\right)^{3/2}}\Delta_{2},
\\
\frac{d\Delta_{2}}{d\ell}&=&\mathcal{H}_{6}(\alpha),
\\
\frac{d\alpha}{d\ell} &=& \left[\frac{1+t\lambda}{\left(1 -
t^{2}\right)^{3/2}}\Delta_{2} - \frac{1}{2}
\left(\mathcal{H}_{1}(\alpha)+\mathcal{H}_{2}(\alpha)
\right)\right]\alpha. \nonumber \\
\end{eqnarray}

\begin{figure}[htbp]
\center
\includegraphics[width=3.35in]{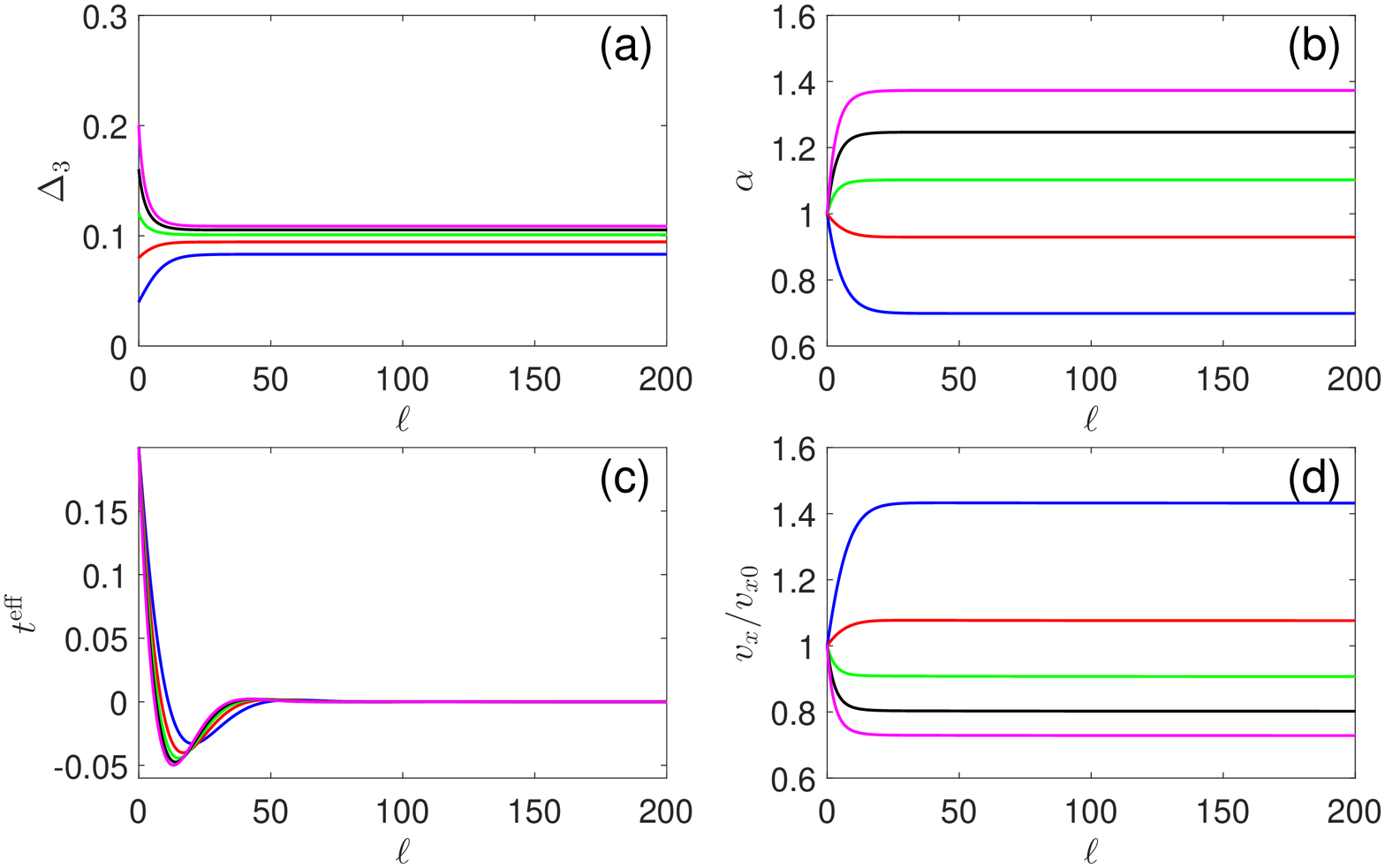}
\caption{Flows of $\Delta_{3}$, $\lambda$, $t^{\mathrm{eff}}$, and
$v_{x}$ due to the interplay between RM and Coulomb interaction.
Blue, red, green, black, magenta lines correspond to the initial
values $\Delta_{3, 0}=0.04, 0.08, 0.12, 0.16, 0.2$, respectively.
Here, $t_{0}=0.2$, $v_{y0}/v_{x0}=1$, and $N=4$.}
\label{Fig:VRGIniRMLRC}
\end{figure}

The $\ell$-dependence of $\Delta_{2}$, $\alpha$, $t^{\mathrm{eff}}$,
and $v_{x}$ can be found in Figs.~\ref{Fig:VRGIniRVPyLRC}(a)-(d). As
$\ell$ grows, $\Delta_{2}$ and $\alpha$ approach to finite values
$\Delta_{2}^{*}$ and $\alpha^{*}$, respectively. Thus the system is
always in the stable quantum critical state characterized by
$\Delta_{2}^{*}$ and $\alpha^{*}$. Coulomb interaction and $y$-RVP
are both marginal. Fig.~\ref{Fig:VRGIniRVPyLRC}(c) shows that
$t^{\mathrm{eff}} \rightarrow 0$, thus the tilt is suppressed. We
see from Fig.~\ref{Fig:VRGIniRVPyLRC}(d) that, $v_{x}$ flows to a
finite values $v_{x}^{*}$. $v_{y}$ also approaches to a finite value
$v_{y}^{*}$, which is not shown here.

The $\ell$-dependence of fermion velocities indicates that the
dynamical exponent recovers the value $z=1$. But the fermion field
acquires a finite anomalous dimension
\begin{eqnarray}
\eta_{\psi} = \frac{\Delta_{2}^{*}}{2\pi v_{x}^{*}v_{y}^{*}}.
\end{eqnarray}
Based on these results, we get the fermion DOS
\begin{eqnarray}
\rho(\omega)\sim\omega^{1+\eta_{\psi}}.
\end{eqnarray}
The specific heat and compressibility exhibit the same behaviors as
the non-interacting case, namely $C_{v}(T)\sim T^{2}$ and
$\kappa(T)\sim T$. As shown in
Appendix~\ref{App:ObservableQuantitiesSQCS}, although the anomalous
dimension $\eta_{\psi}$ is nonzero, it only modifies the
coefficients, leaving the $T$-dependence unchanged.

\begin{table*}[htbp]
\caption{Summary of the flow behaviors of $t$, $\lambda$,
$t^{\mathrm{eff}}$, $\alpha$, and disorder strength parameters in
the low-energy region. The critical scale $\ell_c$ is always finite,
but its precise value is case dependent.
\label{Table:SummaryRGResult}}
\begin{center}
\begin{tabular}{|c|c|c|c|c|c|c|}
\hline\hline    Initial Condition & $\ell$ &$t$ & $\lambda$ &
$t^{\mathrm{eff}}=\frac{t+\lambda}{1+t\lambda}$ & Disorder strength
& $\alpha$
\\
\hline           $\Delta_{0,0}>0$ & when $\ell\rightarrow\ell_{c}$ &
$t=t_{0}$  & $\lambda\rightarrow1$ & $t^{\mathrm{eff}}\rightarrow1$
&
\tabincell{c}{$\Delta_{0}\rightarrow\infty$ \\
$\Delta_{-}\rightarrow\infty$ \\
$\Delta_{-}/\Delta_{0} \rightarrow \infty$} & -
\\
\hline           $\Delta_{1,0}>0$ & when $\ell\rightarrow\ell_{c}$ &
$t=t_{0}$  & $\lambda\rightarrow1$ & $t^{\mathrm{eff}}\rightarrow1$
&
\tabincell{c}{$\Delta_{1}\rightarrow\infty$ \\
$\Delta_{-}\rightarrow\infty$ \\
$\Delta_{-}/\Delta_{1} \rightarrow \infty$} & -
\\
\hline           $\Delta_{2,0}>0$ & when $\ell\rightarrow\infty$ &
$t=t_{0}$  & $\lambda\rightarrow-t_{0}$ &
$t^{\mathrm{eff}}\rightarrow0$ & $\Delta_{2}=\Delta_{2,0}$ & -
\\
\hline           $\Delta_{3,0}>0$ & when $\ell\rightarrow\infty$ &
$t=t_{0}$  & $\lambda\rightarrow-t_{0}$ &
$t^{\mathrm{eff}}\rightarrow0$ & $\Delta_{3}\rightarrow0$ & -
\\
\hline           $\alpha_{0}>0$ & when $\ell\rightarrow\infty$ &
$t\rightarrow0$  & $\lambda=0$ & $t^{\mathrm{eff}}\rightarrow0$ & -
& $\alpha\rightarrow0$
\\
\hline         $\alpha_{0}>0$,  $\Delta_{0,0}>0$   & when
$\ell\rightarrow\ell_{c}$ & $t\rightarrow t^{*}$  &
$\lambda\rightarrow1$ & $t^{\mathrm{eff}}\rightarrow 1$ &
\tabincell{c}{$\Delta_{0}\rightarrow\infty$ \\
$\Delta_{-}\rightarrow\infty$ \\
$\Delta_{-}/\Delta_{0} \rightarrow \infty$} &
\tabincell{c}{$\alpha\rightarrow\infty$\\
$\alpha/\Delta_{-}\rightarrow 0$}
\\
\hline        $\alpha_{0}>0$,  $\Delta_{1,0}>0$   & when
$\ell\rightarrow\ell_{c}$ & $t\rightarrow t^{*}$  &
$\lambda\rightarrow1$ & $t^{\mathrm{eff}}\rightarrow 1$ &
\tabincell{c}{$\Delta_{1}\rightarrow\infty$ \\
$\Delta_{-}\rightarrow\infty$ \\
$\Delta_{-}/\Delta_{1} \rightarrow \infty$} &
\tabincell{c}{$\alpha\rightarrow\infty$\\
$\alpha/\Delta_{-}\rightarrow0$}
\\
\hline          $\alpha_{0}>0$, $\Delta_{2,0}>0$ & when
$\ell\rightarrow\infty$ &$t\rightarrow 0$ & $\lambda\rightarrow 0$ &
$t^{\mathrm{eff}}\rightarrow 0$ & $\Delta_{2}\rightarrow
\Delta_{2}^{*}$ & $\alpha\rightarrow\alpha^{*}$
\\
\hline          $\alpha_{0}>0$, $\Delta_{3,0}>0$ & when
$\ell\rightarrow\infty$ &$t\rightarrow 0$ & $\lambda\rightarrow 0$ &
$t^{\mathrm{eff}}\rightarrow 0$ & $\Delta_{3}\rightarrow
\Delta_{3}^{*}$ & $\alpha\rightarrow \alpha^{*}$
\\
\hline \hline
\end{tabular}
\end{center}
\end{table*}

\subsubsection{Coulomb interaction and RM}

Coulomb interaction and RM give rise to
\begin{eqnarray}
\frac{d v_{x}}{d \ell}&=& \left[\mathcal{H}_{1}(\alpha)
-\frac{1+t\lambda}{\left(1-t^{2}\right)^{3/2}}\Delta_{3}\right]v_{x},
\\
\frac{d v_{y}}{d\ell} &=& \left[\mathcal{H}_{2}(\alpha) -
\frac{1+t\lambda}{\left(1-t^{2}\right)^{3/2}}\Delta_{3}\right]
v_{y}, \\
\frac{d(v_{y}/v_{x})}{d\ell}&=&\left[\mathcal{H}_{2}(\alpha) -
\mathcal{H}_{1}(\alpha)\right] \frac{v_{y}}{v_{x}},
\\
\frac{d(tv_{x})}{d \ell}&=&\left[-\frac{1+t\lambda}{\left(1 -
t^{2}\right)^{3/2}}\Delta_{3}t + \mathcal{H}_{3}(\alpha)
\right]v_{x},
\\
\frac{d t}{d\ell} &=& \mathcal{H}_{3}(\alpha) -
\mathcal{H}_{1}(\alpha)t,
\\
\frac{d\Delta_{3}}{d\ell} &=& \left[-2\frac{1+t\lambda}{\left(1 -
t^{2}\right)^{3/2}}\Delta_{3}+\mathcal{H}_{7}(\alpha)\right]
\Delta_{3}, \\
\frac{d\alpha}{d\ell}&=&\left[\frac{1+t\lambda}{\left(1-t^{2}
\right)^{3/2}}\Delta_{3} - \frac{1}{2}\left(\mathcal{H}_{1}(\alpha)
+ \mathcal{H}_{2}(\alpha)\right)\right]\alpha. \nonumber \\
\end{eqnarray}
As illustrated by Figs.~\ref{Fig:VRGIniRMLRC}(a) and (b),
$\Delta_{3}\rightarrow\Delta_{3}^{*}$ and
$\alpha\rightarrow\alpha^{*}$, where $\Delta_{3}^{*}$ and
$\alpha^{*}$ are two constants. The effective tilt parameter
$t^{\mathrm{eff}} \rightarrow 0$, and $v_{x}\rightarrow v_{x}^{*}$
in the lowest energy limit, which can be easily seen from
Figs.~\ref{Fig:VRGIniRMLRC}(c) and (d).

The system also flows to a stable quantum critical state in which
the dynamical exponent $z=1$ and the fermion anomalous dimension
\begin{eqnarray}
\eta_{\psi} = \frac{\Delta_{3}^{*}}{2\pi v_{x}^{*}v_{y}^{*}}.
\end{eqnarray}
These results are qualitatively very similar to those induced by the
interplay between Coulomb interaction and $y$-RVP. Once again, the
DOS $\rho(\omega)\sim\omega^{1+\eta_{\psi}}$, the specific heat
$C_{v}(T)\sim T^{2}$, and the compressibility $\kappa(T)\sim T$.

\section{Summary and Discussion \label{Sec:Summary}}

In summary, we have studied the physical effects of Dirac cone tilt
on the low-energy behaviors of 2D DSM by performing a RG analysis of
the interplay between Coulomb interaction and quenched disorder. For
the tilt along $x$-axis, there are generically four types of
disorder: RSP, $x$-RVP, $y$-RVP, and RM. We find that RSP and
$x$-RVP are distinct from $y$-RVP and RM. As long as the tilt is
finite, RSP cannot exist on its own and its coupling to the Dirac
fermions inevitably generates a new type of disorder. The
dynamically generated disorder plays the dominant role in the
low-energy region, and drives an SM-to-CDM quantum phase transition.
As the result, the fermions acquire a finite disorder scattering
rate. Moreover, the originally isolated Dirac points are replaced by
a bulk Fermi arc. We also find that $x$-RVP leads to nearly the same
low-energy behaviors as RSP. These results are not altered when the
Coulomb interaction is incorporated. Different from RSP and $x$-RVP,
$y$-RVP or RM can exist alone without generating other types of
disorder. In addition, both $y$-RVP and RM tend to suppress the
Dirac cone tilt. When the Coulomb interaction and $y$-RVP (or RM)
exist concomitantly, they cooperate to produce a stable quantum
critical state, in which the dynamical exponent $z=1$ and the
fermion anomalous dimension is nonzero. All these results are
summarized in Table~\ref{Table:SummaryRGResult}. To characterize the
low-energy behaviors, we also calculate the fermion DOS, specific
heat, and compressibility in various conditions, and summarize the
results in Table~\ref{Table:SummaryObQuant}.

\begin{table*}[htbp]
\caption{Summary of low-energy or low-temperature properties of some
observable quantities, including DOS, specific heat, and
compressibility, obtained in different conditions.
\label{Table:SummaryObQuant}}
\begin{center}
\begin{tabular}{|c|c|c|c|c|}
\hline\hline    Initial Condition & DOS $\rho(\omega)$ & Specific
heat $C_{v}(T)$ & Compressibility $\kappa(T)$
\\
\hline          Clean and Free & $\rho(\omega)\sim\omega$ &
$C_{v}(T)\sim T^{2}$ & $\kappa(T)\sim T$
\\
\hline           $\Delta_{0,0}>0$ & $\rho(0)>0$  &
$C_{v}(T)\sim\rho(0)T$ & $\kappa(0)>0$
\\
\hline           $\Delta_{1,0}>0$ & $\rho(0)>0$  &
$C_{v}(T)\sim\rho(0)T$ & $\kappa(0)>0$
\\
\hline           $\Delta_{2, 0}>0$ &
$\rho(\omega)\sim\omega^{(1-\eta_{0})/(1+\eta_{0})}$  &
$C_{v}(T)\sim T^{2/(1+\eta_{0})}$ & $\kappa(T)\sim
T^{(1-\eta_{0})/(1+\eta_{0})}$
\\
\hline          $\Delta_{3,0}>0$ &
$\rho(\omega)\sim\omega\ln\left(\omega_{0}/\omega\right)$  &
$C_{v}(T)\sim T^{2}\ln(T_{0}/T)$ & $\kappa(T)\sim T\ln(T_{0}/T)$
\\
\hline          $\alpha_{0}>0$ &
$\rho(\omega)\sim\omega/\ln^{2}(\omega_{0}/\omega)$  & $C_{v}(T)\sim
T^{2}/\ln^{2}(T_{0}/T)$ & $\kappa(T)\sim T/\ln^{2}(T_{0}/T)$
\\
\hline           $\alpha_{0}>0$, $\Delta_{0,0}>0$ & $\rho(0)>0$  &
$C_{v}(T)\sim\rho(0)T$ & $\kappa(0)>0$
\\
\hline           $\alpha_{0}>0$, $\Delta_{1,0}>0$ & $\rho(0)>0$  &
$C_{v}(T)\sim\rho(0)T$ & $\kappa(0)>0$
\\
\hline          $\alpha_{0}>0$, $\Delta_{2,0}>0$ &
$\rho(\omega)\sim\omega^{1+\eta_{\psi}(\Delta_{2}^{*})}$  &
$C_{v}(T)\sim T^{2}$  & $\kappa(T)\sim T$
\\
\hline           $\alpha_{0}>0$, $\Delta_{3,0}>0$ &
$\rho(\omega)\sim\omega^{1+\eta_{\psi}(\Delta_{3}^{*})}$  &
$C_{v}(T)\sim T^{2}$ & $\kappa(T)\sim T$
\\
\hline \hline
\end{tabular}
\end{center}
\end{table*}

It is useful to highlight the unusual effects caused by the tilt.
For untilted 2D DSM, previous studies \cite{Ludwig94, Ostrovsky06,
Evers08, Foster12} have already confirmed that any of the four types
of disorder can individually exist. More concretely, RSP is relevant
and converts the DSM into a CDM, in which the fermions have finite
disorder scattering rate but no bulk Fermi arc appears. The two
components of RVP, namely $x$-RVP and $y$-RVP, are equivalent: they
are marginal and lead to stable quantum critical state characterized
by power-law corrections to observable quantities. RM is marginally
irrelevant and merely causes weak logarithmic-like corrections to
observable quantities. When the Dirac cone is tilted along $x$-axis,
$x$-RVP becomes entirely different from $y$-RVP. In fact, $x$-RVP
gives rise to nearly the same physical consequences as RSP: they
always dynamically generate a new type of disorder, and induce a
bulk Fermi arc. In the case of zero tilt, these two features are
both absent. In contrast, $y$-RVP or RM leads to nearly the same
low-energy properties in 2D tilted DSM as those of the untilted
case.

The interplay of Coulomb interaction and disorder in untilted 2D DSM
has also been studied extensively \cite{Ye98, Ye99, Stauber05,
Herbut08, Vafek08, Foster08, WangLiu14}. If Coulomb interaction and
RVP are both considered, 2D DSM is driven to enter into a stable
quantum critical state, in which the fermion field acquires a finite
anomalous dimension but the dynamical exponent becomes $z=1$
\cite{Ye98, Ye99, Stauber05, Herbut08, Vafek08, Foster08,
WangLiu14}. Coexistence of Coulomb interaction and RM leads to
similar stable quantum critical state \cite{Ye98, Ye99, Stauber05,
Foster08, WangLiu14}. The behaviors induced by $y$-RVP and RM are
not qualitatively altered by the tilt. Actually, tilt mainly changes
the physical effects of RSP and $x$-RVP.

Sikkenk and Fritz \cite{Sikkenk17} studied the disorder effects on
3D WSM tilted along a generic direction. They found that RSP and RVP
can exist individually in untilted system but always generate each
other at finite tilt. This is similar to the result obtained in this
work. For weak RSP and RVP, their strength parameters both flow to
zero in the lowest energy limit. However, both RSP and RVP flow to
the strong coupling regime if the initial strength is large enough,
which generates a finite scattering rate. In the latter case, there
should also emerge a bulk Fermi arc, although this conclusion was
not noticed in Ref.~\cite{Sikkenk17}. In Ref.~\cite{Detassis17},
Detassis \emph{et al.} considered the effects caused by the Coulomb
interaction on tilted 3D DSM/WSM, and showed that the tilt is
completely suppressed by the Coulomb interaction, consistent with
the result obtain in the context of tilted 2D DSM \cite{Isobe12,
LeeLee18}. Pozo \emph{et al.} \cite{Pozo18} analyzed the influence
of the electromagnetic field on 3D DSM/WSM with finite tilt, and
found that the tilt parameter approaches to a finite value in the
lowest energy limit when the polarization of photon is properly
taken into account.

Recently, Papaj and Fu \cite{Papaj18} investigated disorder effects
in a model in which the Dirac fermions come from two distinct
orbitals. In this model, disorder acts on two orbitals differently.
They showed \cite{Papaj18} that a finite tilt is generated naturally
due to the orbit-dependent disorder scattering even when the Dirac
cone is initially not tilted. Because two orbitals acquire different
disorder scattering rates, the original Dirac points are replaced by
a bulk Fermi arc \cite{Papaj18}. Later, Zhao \emph{et al.}
\cite{Zhao18} extended the analysis of Papaj and Fu \cite{Papaj18}
to the more generic case in which several types of disorder coexist,
and obtained a condition for the Fermi arc to emerge. The model
considered in our work differs from the one studied in
Refs.~\cite{Papaj18, Zhao18} in that the two components of the
spinor field have different physical origin. According to our
results, both RSP and $x$-RVP can dynamically generate a new type of
disorder, which then plays an overwhelming role at low energies. The
striking phenomenon of dynamical disorder generation was not
considered in Refs.~\cite{Papaj18, Zhao18}.

\section*{ACKNOWLEDGEMENTS}

We would thank Peng-Lu Zhao for helpful discussions. We acknowledge
the support by the National Natural Science Foundation of China
under Grants 11574285 and 11504379. Z.K.Y. and G.Z.L. are partly
supported by the Fundamental Research Funds for the Central
Universities (P. R. China) under Grant WK2030040085. J.R.W. is also
supported by the Natural Science Foundation of Anhui Province under
Grant 1608085MA19.

\appendix

\section{Deriving RG equations \label{App:DerivationRG}}

Here we present the detailed derivation of the coupled RG equations
for all the involved model parameters.

\subsection{Self-energy corrections of fermions}

The fermion self-energy stems from two interactions: Coulomb
interaction and disorder scattering. We consider two cases in order.

\subsubsection{Self-energy induced by Coulomb interaction}

The self-energy of fermions induced by the long-range Coulomb
interaction is defined as
\begin{eqnarray}
\Sigma^{C}(i\omega,\mathbf{k}) &=& -\int'\frac{d\Omega}{2\pi}
\frac{d^2\mathbf{q}}{(2\pi)^{2}} G_{0}\left(i\omega+i\Omega,
\mathbf{k}+\mathbf{q}\right)\nonumber
\\
&&\times D(\mathbf{q}),
\label{Eq:FermionSelfEnergyCoulomb}
\end{eqnarray}
where $\int'$ means that the integration requires a proper choice of
the momentum shell. We choose to integrate in the ranges of $-\infty
< \Omega < \infty$ and $b\Lambda < E_{\mathbf{k}} < \Lambda$, where
$E_{\mathbf{k}} = tv_{x}k_{x} + \sqrt{v_{x}^{2}k_{x}^{2} +
v_{y}^{2}k_{y}^{2}}$ and $b = e^{-\ell}$. Substituting
Eqs.~(\ref{Eq:DressedCoulomb}) and (\ref{Eq:FermionPropagatorNew})
into Eq.~(\ref{Eq:FermionSelfEnergyCoulomb}) and expanding up to the
leading order, we obtain
\begin{eqnarray}
\Sigma^{C}(i\omega,\mathbf{k}) \approx \left[\Sigma^C_{t}v_{x}k_{x} +
\Sigma^C_{x}v_{x}k_{x}\sigma_{x} +
\Sigma^C_{y}v_{y}k_{y}\sigma_{y}\right]\ell, %\nonumber \\
\end{eqnarray}
where
\begin{eqnarray}
\Sigma^C_{t}&=&\lambda\left(1+t\lambda\right)f_{A}(\alpha),
\label{Eq:SigmaCw} \\
\Sigma^C_{x}&=&\left(1+t\lambda\right)f_{A}(\alpha),
\label{Eq:SigmaCx} \\
\Sigma^C_{y}&=&\left(1+t\lambda\right)^{2}f_{B}(\alpha),
\label{Eq:SigmaCy}
\end{eqnarray}
where $f_{A}$ and $f_{B}$ are given by Eqs.~(\ref{Eq:fA}) and
(\ref{Eq:fB}).

\begin{figure}[htbp]
\center
\includegraphics[width=2.6in]{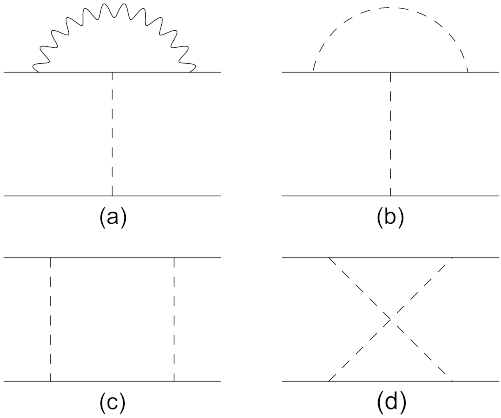}
\caption{Diagrams of vertex corrections to fermion-disorder
coupling. Solid, dashed, and wave lines represent fermion
propagator, disorder scattering, and Coulomb interaction.
\label{Fig:VertexCorrection}}
\end{figure}

\subsubsection{Self-energy induced by disorder scattering}

The fermion self-energy caused by disorder scattering is given by
\begin{eqnarray}
\Sigma^{\mathrm{dis}}(i\omega) = \sum_{n}\Delta_{n} \int'
\frac{d^2\mathbf{p}}{(2\pi)^{2}} \Gamma_{n}G_{0}
\left(i\omega,\mathbf{p}\right)\Gamma_{n},
\label{Eq:FermionSelfEnergyDisorder}
\end{eqnarray}
where $\sum_{n}\equiv\sum_{n=0,1,2,3,-}$. Substituting
Eq.~(\ref{Eq:FermionPropagatorNew}) into
Eq.~(\ref{Eq:FermionSelfEnergyDisorder}) and retaining the leading
order contribution, we get
\begin{eqnarray}
\Sigma^{\mathrm{dis}}(i\omega) &\approx& \left(i\omega
\Sigma^{\mathrm{dis}}_{\omega}\sigma_{0} + i\omega
\Sigma^{\mathrm{dis}}_{\lambda} \sigma_{x}\right)\ell,
\end{eqnarray}
where
\begin{eqnarray}
\Sigma^{\mathrm{dis}}_{\omega} &=& -\left[\left(\Delta_{0} +
\Delta_{1}+\Delta_{2}+\Delta_{3}\right) + (1+t)
\frac{\Delta_{-}}{2}\right]\nonumber
\\
&&\times\frac{1+t\lambda}{2\pi v_{x}v_{y}
\left(1-t^{2}\right)^{3/2}}, \label{Eq:SigmaOmega} \\
\Sigma^{\mathrm{dis}}_{\lambda} &=& \left[t\left(\Delta_{0}+\Delta_{1} -
\Delta_{2}-\Delta_{3}\right)+\left(1+t\right)
\frac{\Delta_{-}}{2}\right]\nonumber
\\
&&\times \frac{1+t\lambda}{2\pi v_{x}v_{y} \left(1-t^{2}
\right)^{3/2}}. \label{Eq:SigmaLambda}
\end{eqnarray}

\subsection{Corrections to fermion-disorder coupling}

Diagram \ref{Fig:VertexCorrection}(a) represents the correction to
the fermion-disorder coupling induced by Coulomb interaction. We
have
\begin{eqnarray}
W^{a} &=& \sum_{n}W_{n}^{a}, \\
W_{n}^{a} &=& -\Delta_{n}\int'\frac{d\Omega}{2\pi}
\frac{d^2\mathbf{q}}{(2\pi)^{2}}\left(\psi^{\dag}
G_{0}\left(i\Omega,\mathbf{q}\right)\Gamma_{n}
G_{0}\left(i\Omega,\mathbf{q}\right)\psi\right) \nonumber
\\
&&\times\left(\psi^{\dag}\Gamma_{n}\psi\right)D(\mathbf{q}).
\end{eqnarray}
After tedious but straightforward calculations, we finally obtain
\begin{eqnarray}
W^{a} = \left[\sum_{n}\delta\Delta_{n}^{a}
\left(\psi^{\dag}\Gamma_{n}\psi\right)
\left(\psi^{\dag}\Gamma_{n}\psi\right)\right]\ell,
\end{eqnarray}
where
\begin{eqnarray}
\delta\Delta_{0}^{a}&=&\left[\lambda\left(1+2\lambda\right)
\Delta_{0}+\lambda \Delta_{1}-\frac{1}{4}
\left(1-\lambda^{2}\right)\Delta_{-}\right]\nonumber
\\
&&\times\frac{f_{A}(\alpha)}{2\left(1-\lambda^{2}\right)^{3/2}},
\label{Eq:deltaDelta0a}
\\
\delta\Delta_{1}^{a} &=& \left[\lambda\Delta_{0} +
\left(2+\lambda\right)\Delta_{1} + \frac{1}{4}
\left(1-\lambda^{2}\right)\Delta_{-}\right]\nonumber
\\
&&\times\frac{f_{A}(\alpha)}{2\left(1 - \lambda^{2}\right)^{3/2}},
\label{Eq:deltaDelta1a}
\\
\delta\Delta_{2}^{a}&=&\left(1+t\lambda\right)^{2}\Delta_{2}
\frac{f_{B}(\alpha)}{\left(1-\lambda^{2}\right)^{3/2}},
\label{Eq:deltaDelta2a}
\\
\delta\Delta_{3}^{a}&=&\left[\left(1-\lambda^{2}\right)f_{A}(\alpha)
+\left(1+t\lambda\right)^{2}f_{B}(\alpha)\right]\nonumber
\\
&&\times\frac{\Delta_{3}}{\left(1-\lambda^{2}\right)^{3/2}},
\label{Eq:deltaDelta3a}
\\
\delta\Delta_{-}^{a}&=&\left[-4\lambda\left(\Delta_{0}+\Delta_{1}\right)
+\left(1+\lambda\right)^{2}\Delta_{-}\right]\nonumber
\\
&&\times\frac{f_{A}(\alpha)}
{2\left(1-\lambda^{2}\right)^{3/2}}. \label{Eq:deltaDeltaMinusa}
\end{eqnarray}
In the derivation of these equations, we have encountered a new
coupling term $\left(\psi^{\dag}\sigma_{x}\psi\right)
\left(\psi^{\dag}\sigma_{0}\psi\right)$, which does not exist in the
starting action but is generated by fermion-disorder interaction. To
deal with this new term, we find it convenient to decouple it as
follows
\begin{eqnarray}
\left(\psi^{\dag}\sigma_{x}\psi\right)
\left(\psi^{\dag}\sigma_{0}\psi\right) &=&
-2\left(\psi^{\dag}\sigma_{-}\psi\right)
\left(\psi^{\dag}\sigma_{-}\psi\right)\nonumber
\\
&&+\frac{1}{2}\left(\psi^{\dag}\sigma_{0}\psi\right)
\left(\psi^{\dag}\sigma_{0}\psi\right)\nonumber
\\
&&+\frac{1}{2}\left(\psi^{\dag}\sigma_{x}\psi\right)
\left(\psi^{\dag}\sigma_{x}\psi\right). \label{Eq:NewCouplingRelation}
\end{eqnarray}
Thus the dynamically generated term is actually a combination of
RSP, $x$-RVP, and a new type of disorder defined by the matrix
$\sigma_{-}$. Relation (\ref{Eq:NewCouplingRelation}) is also
employed in the following if the term
$\left(\psi^{\dag}\sigma_{x}\psi\right)
\left(\psi^{\dag}\sigma_{0}\psi\right)$ appears.

The contribution for the fermion-disorder coupling from the Feynman
diagram \ref{Fig:VertexCorrection}(b) is given by
\begin{eqnarray}
W^{b} &=& \sum_{n}W_{n}^{b},  \label{Eq:WbAll}
\\
W_{n}^{b}&=&\sum_{m}\int'\frac{d^2\mathbf{p}}{(2\pi)^{2}}
\left(\psi^{\dag}\sigma_{m}G_{0}(0,\mathbf{p})\sigma_{n}
G_{0}\left(0,\mathbf{p}\right)\sigma_{m}\psi\right)\nonumber
\\
&&\times\left(\psi^{\dag}\sigma_{n}\psi\right). \label{Eq:WbComponent}
\end{eqnarray}
Substituting Eq.~(\ref{Eq:FermionPropagatorNew}) into
Eqs.~(\ref{Eq:WbAll}) and (\ref{Eq:WbComponent}) and performing the
calculation directly, we obtain
\begin{eqnarray}
W^{b}&=&\left[\sum_{n}\delta\Delta_{n}^{b}
\left(\psi^{\dag}\Gamma_{n}\psi\right)
\left(\psi^{\dag}\Gamma_{n}\psi\right)\right]\ell,
\end{eqnarray}
where
\begin{widetext}
\begin{eqnarray}
\delta\Delta_{0}^{b} &=& \Bigg[\left(1-\frac{t}{2}\right)
\Delta_{0}^{2} + \left(1-t\right)\Delta_{0}\Delta_{1} +
\left(1+\frac{t}{2}\right) \Delta_{0}\Delta_{2} + \left(1 +
\frac{t}{2}\right)\Delta_{0}\Delta_{3}-\frac{t}{2}
\Delta_{1}^{2}-\frac{t}{2}\Delta_{1}\Delta_{2} -
\frac{t}{2}\Delta_{1}\Delta_{3} \nonumber
\\
&&+\left(\frac{3}{8}+\frac{t}{4}-\frac{t^{2}}{8}\right)
\Delta_{0}\Delta_{-} + \left(\frac{1}{8}-\frac{t}{4} -
\frac{3t^{2}}{8}\right)\Delta_{1}\Delta_{-} +
\frac{\left(1+t\right)^{2}}{8}\left(\Delta_{2} +
\Delta_{3}\right)\Delta_{-}\Bigg] \frac{1}{2\pi
v_{x}v_{y}\left(1-t^{2}\right)^{3/2}}, \nonumber\\ \label{Eq:deltaDelta0b}
\\
\delta\Delta_{1}^{b}&=&\Bigg[-\frac{t}{2}\Delta_{0}^{2} -
t\left(1-t\right)\Delta_{0}\Delta_{1} + \frac{t}{2}\Delta_{0}
\Delta_{2}+\frac{t}{2}\Delta_{0}\Delta_{3} - t\left(\frac{1}{2} -
t\right)\Delta_{1}^{2}-t\left(\frac{1}{2}+t\right)\Delta_{1}
\left(\Delta_{2}+\Delta_{3}\right)\nonumber
\\
&&-\left(\frac{3}{8}+\frac{t}{4}-\frac{t^{2}}{8}\right)\Delta_{0}\Delta_{-}
-\left(\frac{1}{8}-\frac{t}{4}-\frac{3t^{2}}{8}\right)\Delta_{1}\Delta_{-}
-\frac{\left(1+t\right)^{2}}{8}\left(\Delta_{2}+\Delta_{3}\right)\Delta_{-}\Bigg]
\frac{1}{2\pi v_{x}v_{y}\left(1-t^{2}\right)^{3/2}}, \nonumber\\
\label{Eq:deltaDelta1b}
\\
\delta\Delta_{2}^{b}&=&0, \label{Eq:deltaDelta2b}
\\
\delta\Delta_{3}^{b}&=&-\left(1-t^{2}\right)\left(\Delta_{0}-\Delta_{1}
-\Delta_{2}+\Delta_{3}\right)\frac{1}{2\pi
v_{x}v_{y}\left(1-t^{2}\right)^{3/2}}, \label{Eq:deltaDelta3b}
\\
\delta\Delta_{-}^{b}&=&\Bigg[2t\left(\Delta_{0}^{2}+2\Delta_{0}\Delta_{1}
-\Delta_{0}\Delta_{2}-\Delta_{0}\Delta_{3}+\Delta_{1}^{2}
+\Delta_{1}\Delta_{2}+\Delta_{1}\Delta_{3}\right)
+\frac{1}{2}\left(3+4t+t^{2}\right)\Delta_{0}\Delta_{-}\nonumber
\\
&&+\frac{1}{2}\left(1+4t+3t^{2}\right)\Delta_{1}\Delta_{-} +
\frac{1-t^{2}}{2}\left(\Delta_{2}+\Delta_{3}\right)\Delta_{-} +
\frac{\left(1+t\right)^{2}}{2}\Delta_{-}^{2}\Bigg]\frac{1}{2\pi
v_{x}v_{y}\left(1-t^{2}\right)^{3/2}}. \label{Eq:deltaDeltaMinusb}
\end{eqnarray}

Diagram \ref{Fig:VertexCorrection} (c) and diagram
\ref{Fig:VertexCorrection} (d) give rise to the following
contributions to the fermion-disorder coupling vertex
\begin{eqnarray}
W^{c+d} &=& \sum_{n}\sum_{m\le n}W_{mn}^{c+d}, \label{Eq:WcdAll}
\\
W_{mn}^{c+d} &=& \Delta_{m}\Delta_{n}\int'
\frac{d^2\mathbf{p}}{(2\pi)^{2}} \left(\psi^{\dag}
\Gamma_{m}G_{0}(0,\mathbf{p})\Gamma_{n}(0,\mathbf{p})\psi\right)
\left(\psi^{\dag}\left(\Gamma_{n}G_{0}(0,\mathbf{p}) \Gamma_{m} +
\Gamma_{m}G_{0}(0,-\mathbf{p})\Gamma_{n}\right)\psi\right).
\label{Eq:WcdComponent}
\end{eqnarray}
Substituting Eq.~(\ref{Eq:FermionPropagatorNew}) into
Eqs.~(\ref{Eq:WcdAll}) and (\ref{Eq:WcdComponent}), we get
\begin{eqnarray}
W^{c+d} = \left[\sum_{n}\delta\Delta_{n}^{c+d}
\left(\psi^{\dag}\Gamma_{n}\psi\right) \left(\psi^{\dag}
\Gamma_{n}\psi\right)\right]\ell, \\
\end{eqnarray}
where
\begin{eqnarray}
\delta\Delta_{0}^{c+d} &=& \left[\left(1-t^{2}\right)\Delta_{1} +
\left(1+t\right)\Delta_{2}\right] \frac{\Delta_{3}}{\pi
v_{x}v_{y}\left(1-t^{2}\right)^{3/2}}, \label{Eq:deltaDelta0cd}
\\
\delta\Delta_{1}^{c+d}&=&\left[\left(1-t^{2}\right)\Delta_{0} +
t\left(1+t\right)\Delta_{2}\right] \frac{\Delta_{3}}{\pi
v_{x}v_{y}\left(1-t^{2}\right)^{3/2}}, \label{Eq:deltaDelta1cd}
\\
\delta\Delta_{2}^{c+d}&=& \left[\left(\Delta_{0} + t^{2}
\Delta_{1}\right) + \frac{\left(1+t\right)^{2} \Delta_{-}}{4}\right]
\frac{\Delta_3}{\pi v_{x}v_{y}
\left(1-t^{2}\right)^{3/2}}, \label{Eq:deltaDelta2cd}
\\
\delta\Delta_{3}^{c+d} &=& \Big[\left(1-t^{2}\right)
\Delta_{0}\Delta_{1} +\Delta_{0}\Delta_{2} +
t^{2}\Delta_{1}\Delta_{2} + \frac{1-t^{2}}{4}
\left(\Delta_{0}+\Delta_{1}\right)\Delta_{-} +
\frac{\left(1+t\right)^{2}}{4} \Delta_{2} \Delta_{+}\Big]
\frac{1}{\pi v_{x}v_{y} \left(1-t^{2}\right)^{3/2}},\nonumber\\
\label{Eq:deltaDelta3cd}
\\
\delta\Delta_{-}^{c+d} &=& \left[-4t\Delta_{2} +
\left(1-t^{2}\right) \Delta_{-}\right]
\frac{\Delta_3}{\pi v_{x}v_{y}
\left(1-t^{2}\right)^{3/2}}. \label{Eq:deltaDeltaMinuscd}
\end{eqnarray}
\end{widetext}

\subsection{RG analysis}

The original action of fermions is
\begin{eqnarray}
S_{f}&=&\int\frac{d\omega}{2\pi}\frac{d^2\mathbf{k}}{(2\pi)^{2}}
\psi^{\dag}(\omega,\mathbf{k})\left[i\omega\sigma_{0}-i\lambda
\omega\sigma_{x}\right.\nonumber \\
&&\left.-\left(t\sigma_{0}+\sigma_{x}\right)v_{x}k_{x} -
v_{y}k_{y}\sigma_{y}\right] \psi(\omega,\mathbf{k}).
\end{eqnarray}
The original action for fermion-disorder coupling has the form
\begin{eqnarray}
S_{dis}&=&\sum_{n}\frac{\Delta_{n}}{2}\int\frac{d\omega_{1}
d\omega_{2}d^2\mathbf{k}_{1}d^2\mathbf{k}_{2}
d^2\mathbf{k}_{3}}{(2\pi)^{8}}\nonumber \\
&&\times\left(\psi^{\dag}(\omega_{1},\mathbf{k}_{1})\Gamma_{n}
\psi(\omega_{1},\mathbf{k}_{2})\right)
\left(\psi^{\dag}(\omega_{2},\mathbf{k}_{3})\Gamma_{n}\right.\nonumber
\\
&&\left.\times\psi(\omega_{2},-\mathbf{k}_{1} -
\mathbf{k}_{2}-\mathbf{k}_{3})\right),
\end{eqnarray}
where $n=0, 1, 2, 3, -$. Including the corrections obtained in the
last subsections leads to
\begin{eqnarray}
S_{f}'&=&\int\frac{d\omega}{2\pi}\frac{d^2\mathbf{k}}{(2\pi)^{2}}
\psi^{\dag}(\omega,\mathbf{k})\left[i\omega\left(1-\Sigma^{\mathrm{dis}}_{\omega}
\ell\right)\sigma_{0}\right.\nonumber
\\
&&-i\lambda\left(1+\frac{\Sigma^{\mathrm{dis}}_{\lambda}}{\lambda}\ell\right)\omega\sigma_{x}
-\left[t\left(1+\frac{\Sigma^C_{t}}{t}\ell\right)\sigma_{0}\right.\nonumber
\\
&&\left.\left.+\left(1+\Sigma^C_{x}\ell\right)\sigma_{x}\right]v_{x}k_{x}-
\left(1+\Sigma^C_{y}\ell\right)v_{y}k_{y}\sigma_{y}\right]\nonumber
\\
&&\times\psi(\omega,\mathbf{k}), \label{Eq:ActionFermionCorrected}
\end{eqnarray}
and
\begin{eqnarray}
S_{dis}'&=&\sum_{n}\frac{\left(\Delta_{n}+\delta\Delta_{n}
\ell\right)}{2}\int\frac{d\omega_{1}
d\omega_{2}d^2\mathbf{k}_{1}d^2\mathbf{k}_{2}
d^2\mathbf{k}_{3}}{(2\pi)^{8}}\nonumber
\\
&&\times\left(\psi^{\dag}(\omega_{1},\mathbf{k}_{1})\Gamma_{n}
\psi(\omega_{1},\mathbf{k}_{2})\right)
\left(\psi^{\dag}(\omega_{2},\mathbf{k}_{3})\Gamma_{n}\right.\nonumber
\\
&&\left.\times\psi(\omega_{2},-\mathbf{k}_{1} -
\mathbf{k}_{2}-\mathbf{k}_{3})\right),
\end{eqnarray}
where
\begin{eqnarray}
\delta\Delta_{n} = 2\left(\delta\Delta_{n}^{a} +
\delta\Delta_{n}^{b}+\delta_{n}^{c+d}\right).
\label{Eq:ActionFermionDisCoupCorrected}
\end{eqnarray}

Making use of the scaling transformations
\begin{eqnarray}
\omega'&=& b^{-1}\omega,
\\
k_{x}'&=&b^{-1}k_{x},
\\
k_{y}'&=&b^{-1}k_{y},
\\
\psi'&=& Z_{\psi}\psi,
\\
v_{x}' &=& Z_x v_{x},
\\
v_{y}'&=& Z_{y}v_{y},
\\
t'&=& Z_{t} t, \\
\lambda'&=& Z_{\lambda}\lambda,
\\
\Delta_{0}'&=& Z_{0}\Delta_{0},
\\
\Delta_{1}'&=& Z_{1}\Delta_{1},
\\
\Delta_{2}'&=& Z_{2}\Delta_{2},
\\
\Delta_{3}'&=& Z_{3}\Delta_{3},
\\
\Delta_{-}'&=&Z_{-}\Delta_{-},
\end{eqnarray}
we find the following identities
\begin{eqnarray}
b^{-4} Z_{\psi}^{2} &=& \left(1-\Sigma^{\mathrm{dis}}_{\omega}
\ell\right), \\
Z_{x}&=&\left(1+\Sigma^{\mathrm{dis}}_{\omega}\ell\right)
\left(1+\Sigma^C_{x}\ell\right), \\
Z_{y}&=&\left(1+\Sigma^{\mathrm{dis}}_{\omega}\ell\right)
\left(1+\Sigma^C_{y}\ell\right),
\\
Z_{t}Z_{x}&=&\left(1+\Sigma^{\mathrm{dis}}_{\omega}\ell\right)
\left(1+\frac{\Sigma^C_{t}}{t}\ell\right),
\\
Z_{\lambda}&=&\left(1+\Sigma^{\mathrm{dis}}_{\omega}\ell\right)
\left(1+\frac{\Sigma^{dis}_{\lambda}}{\lambda}\ell\right),
\\
Z_{0}&=&\left(1+2\Sigma^{\mathrm{dis}}_{\omega}\ell\right)
\left(1+\frac{\delta\Delta_{0}}{\Delta_{0}}\ell\right),
\\
Z_{1}&=&\left(1+2\Sigma^{\mathrm{dis}}_{\omega}\ell\right)
\left(1+\frac{\delta\Delta_{1}}{\Delta_{1}}\ell\right),
\\
Z_{2}&=&\left(1+2\Sigma^{\mathrm{dis}}_{\omega}\ell\right)
\left(1+\frac{\delta\Delta_{2}}{\Delta_{2}}\ell\right),
\\
Z_{3}&=&\left(1+2\Sigma^{\mathrm{dis}}_{\omega}\ell\right)
\left(1+\frac{\delta\Delta_{3}}{\Delta_{3}}\ell\right),
\\
Z_{-}&=&\left(1+2\Sigma^{\mathrm{dis}}_{\omega}\ell\right)
\left(1+\frac{\delta\Delta_{-}}{\Delta_{-}}\ell\right).
\end{eqnarray}
Thus, the RG equation for the corresponding parameters can be
written as
\begin{eqnarray}
\frac{dv_{x}}{d\ell}&=&\left(\Sigma^{\mathrm{dis}}_{\omega}+\Sigma^C_{x}\right)
v_{x}, \label{Eq:AppRGEqvx}
\\
\frac{dv_{y}}{d\ell}&=&\left(\Sigma^{\mathrm{dis}}_{\omega}+\Sigma^C_{y}\right)
v_{y}, \label{Eq:AppRGEqvy} \\
\frac{d(v_{y}/v_{x})}{d\ell} &=& \left(\Sigma^C_{y} -
\Sigma^C_{x}\right)\frac{v_{y}}{v_{x}}, \label{Eq:AppRGEqvRatio}
\\
\frac{d(tv_{x})}{d\ell}&=&\Sigma^{\mathrm{dis}}_{\omega}
tv_{x}+\Sigma^C_{t}v_{x}, \label{Eq:AppRGEqwvx}
\\
\frac{dt}{d\ell}&=&\Sigma^C_{t}-t\Sigma^C_{x}, \label{Eq:AppRGEqw}
\\
\frac{d\lambda}{d\ell} &=& \Sigma^{\mathrm{dis}}_{\omega}\lambda +
\Sigma^{\mathrm{dis}}_{\lambda}, \label{Eq:AppRGEqLambda}
\\
\frac{d\alpha}{d\ell}&=&-\left(\Sigma^{\mathrm{dis}}_{\omega} +
\frac{1}{2}\Sigma^C_{x} + \frac{1}{2}\Sigma^C_{y}\right),
\label{Eq:AppRGEqAlpha}
\\
\frac{d\Delta_{0}}{d\ell}&=&2\Sigma^{\mathrm{dis}}_{\omega}\Delta_{0}
+ \delta\Delta_{0}, \label{Eq:AppRGEqDelta0}
\\
\frac{d\Delta_{1}}{d\ell}&=&2\Sigma^{\mathrm{dis}}_{\omega}\Delta_{1}
+ \delta\Delta_{1}, \label{Eq:AppRGEqDelta1}
\\
\frac{d\Delta_{2}}{d\ell}&=&2\Sigma^{\mathrm{dis}}_{\omega}
\Delta_{2}+\delta\Delta_{2}, \label{Eq:AppRGEqDelta2}
\\
\frac{d\Delta_{3}}{d\ell} &=& 2\Sigma^{\mathrm{dis}}_{\omega}
\Delta_{3} + \delta\Delta_{3}, \label{Eq:AppRGEqDelta3}
\\
\frac{d\Delta_{-}}{d\ell} &=& 2\Sigma^{\mathrm{dis}}_{\omega}
\Delta_{-} + \delta\Delta_{-}. \label{Eq:AppRGEqDeltaPlus}
\end{eqnarray}
It is convenient to adopt the redefinition
\begin{eqnarray}
\frac{\Delta_{n}}{2\pi v_{x}v_{y}}\rightarrow\Delta_{n}.
\end{eqnarray}
The RG equation for new $\Delta_{n}$ is
\begin{eqnarray}
\frac{d\Delta_{n}}{d\ell} = \delta\Delta_{n}-\left(\Sigma^C_{x} +
\Sigma^C_{y}\right)\Delta_{n}. \label{Eq:AppRGEqDeltaNew}
\end{eqnarray}
Substituting Eqs.~(\ref{Eq:SigmaCw})-(\ref{Eq:SigmaCy}),
(\ref{Eq:SigmaOmega}) and (\ref{Eq:SigmaLambda}),
(\ref{Eq:deltaDelta0a})-(\ref{Eq:deltaDeltaMinusa}),
(\ref{Eq:deltaDelta0b})-(\ref{Eq:deltaDeltaMinusb}),
(\ref{Eq:deltaDelta0cd})-(\ref{Eq:deltaDeltaMinuscd}) into
Eqs.~(\ref{Eq:AppRGEqvx})-(\ref{Eq:AppRGEqAlpha}) and
(\ref{Eq:AppRGEqDeltaNew}), we obtain the RG equations as shown in
Eqs.~(\ref{Eq:RGEqvx})-(\ref{Eq:RGEqDeltaPlus}).

\section{Observable quantities \label{App:ObservableQuantities}}

We compute the DOS, specific heat, and compressibility in order.

\subsection{DOS}

For the fermion propagator given by
Eq.~(\ref{Eq:FermionPropagatorNew}), the spectral function can be
written as
\begin{eqnarray}
A(\omega,\mathbf{k})&=&\frac{2}{\pi}\pi\mathrm{sgn}(\omega)\omega
\delta\left[\left(\omega-E_{+}(\mathbf{k})\right)
\left(\omega-E_{-}(\mathbf{k})\right)\right]\nonumber
\\
&=&|\omega|\left[ \frac{\delta\left(\omega - E_{+}(\mathbf{k})
\right)}{\left|E_{+}(\mathbf{k})\right|} + \frac{\delta\left(\omega
- E_{-}(\mathbf{k})\right)}{\left|E_{-}(\mathbf{k})\right|}\right], %\nonumber\\
\end{eqnarray}
where
\begin{eqnarray}
E_{\pm}(\mathbf{k}) = t^{\mathrm{eff}}v_{x}^{\mathrm{eff}}k_{x} \pm
\sqrt{\left(v_{x}^{\mathrm{eff}}\right)^{2}k_{x}^{2} +
\left(v_{y}^{\mathrm{eff}}\right)^{2}k_{y}^{2}}.
\end{eqnarray}
We consider the case $0<t^{\mathrm{eff}}<1$. Thus $E_{+}(\mathbf{k})>0$ and
$E_{-}(\mathbf{k})<0$. The DOS is given by
\begin{eqnarray}
\rho(\omega) &=& N\int\frac{d^2\mathbf{k}}{(2\pi)^{2}}
A(\omega,\mathbf{k})\nonumber \\
&=& N\int\frac{d^2\mathbf{k}}{(2\pi)^{2}}|\omega|\left[\frac{\delta
\left(\omega-E_{+}(\mathbf{k})\right)}{\left|E_{+}(\mathbf{k})
\right|} \right.\nonumber
\\
&&\left.+ \frac{\delta\left(\omega-E_{-}(\mathbf{k})
\right)}{\left|E_{-}(\mathbf{k})\right|}\right].
\end{eqnarray}
If $\omega>0$, we have
\begin{eqnarray}
\rho(\omega) = N\int\frac{d^2\mathbf{k}}{(2\pi)^{2}}|\omega|
\frac{\delta\left(\omega -
E_{+}(\mathbf{k})\right)}{\left|E_{+}(\mathbf{k})\right|}.
\end{eqnarray}
Let
\begin{eqnarray}
E &=& t^{\mathrm{eff}}v_{x}^{\mathrm{eff}}k_{x} +
\sqrt{\left(v_{x}^{\mathrm{eff}}\right)^{2}k_{x}^{2} +
\left(v_{y}^{\mathrm{eff}}\right)^{2}k_{y}^{2}},
\label{Eq:IntTransformationA} \\
\tanh \theta &=& \frac{v_{y}^{\mathrm{eff}}
k_{y}}{v_{x}^{\mathrm{eff}}k_{x}}, \label{Eq:IntTransformationB}
\end{eqnarray}
which are equivalent to
\begin{eqnarray}
k_{x} = \frac{E\cos \theta}{v_{x}^{\mathrm{eff}}
\left(t^{\mathrm{eff}}\cos \theta + 1\right)},
\label{Eq:IntTransformationC}
\\
k_{y} = \frac{E\sin \theta}{v_{y}^{\mathrm{eff}}
\left(t^{\mathrm{eff}}\cos \theta + 1\right)}.
\label{Eq:IntTransformationD}
\end{eqnarray}
The measures of integrations satisfy the relation
\begin{eqnarray}
dk_{x}dk_{y} &=& \left|
\begin{array}{cc}
\frac{\partial k_{x}}{\partial E} & \frac{\partial
k_{x}}{\partial\theta}
\\
\frac{\partial k_{y}}{\partial E} & \frac{\partial
k_{y}}{\partial\theta}
\end{array}
\right|dEd\theta\nonumber
\\
&=&\frac{E}{v_{x}^{\mathrm{eff}}v_{y}^{\mathrm{eff}}
\left(t^{\mathrm{eff}}\cos \theta + 1\right)^{2}}dEd\theta.
\label{Eq:IntTransformationMeasure}
\end{eqnarray}
Employing the transformations shown in
Eqs.~(\ref{Eq:IntTransformationA})-(\ref{Eq:IntTransformationMeasure}),
DOS can be further written as
\begin{eqnarray}
\rho(\omega) &=& \frac{N|\omega|}{4\pi^{2}v_{x}^{\mathrm{eff}}
v_{y}^{\mathrm{eff}}}\int_{0}^{2\pi}d\theta
\frac{1}{\left(t^{\mathrm{eff}}\cos \theta + 1\right)^{2}}\nonumber
\\
&&\times\int_{0}^{+\infty}dE \delta\left(\omega-E\right)\nonumber
\\
&=&\frac{N|\omega|}{2\pi v_{x}^{\mathrm{eff}}v_{y}^{\mathrm{eff}}
\left[1-\left(t^{\mathrm{eff}}\right)^{2}\right]^{3/2}}.
\end{eqnarray}
Similarly, if $\omega<0$, we also obtain
\begin{eqnarray}
\rho(\omega) &=& \frac{N|\omega|}{2\pi v_{x}^{\mathrm{eff}}
v_{y}^{\mathrm{eff}}\left[1-\left(t^{\mathrm{eff}}\right)^{2}\right]^{3/2}}.
\end{eqnarray}

\subsection{Specific heat}

We now study the $T$-dependence of the specific heat. The specific
heat can be directly computed from the free energy, which is given
by
\begin{eqnarray}
F_{f}(T) &=& -NT\sum_{\omega_{n}}\int\frac{d^2\mathbf{k}}{(2\pi)^2}
\left\{\ln\left[\left(\omega_{n}^2 + E_{+}^{2}(\mathbf{k})
\right)^{\frac{1}{2}} \right]\right.\nonumber
\\
&&\left.+\ln\left[\left(\omega_{n}^2 + E_{-}^{2}(\mathbf{k})
\right)^{\frac{1}{2}} \right]\right\}.
\end{eqnarray}
It is easily to verify that $F_{f}(T)$ can be further written as
\begin{eqnarray}
F_{f}(T) &=& -NT\sum_{\omega_{n}}\int\frac{d^2\mathbf{k}}{(2\pi)^2}
\ln\left[\omega_{n}^2+E_{+}^{2}(\mathbf{k})\right]. \nonumber \\
\end{eqnarray}
Summing over $\omega_{n}$ yields
\begin{eqnarray}
F_{f}(T) = -N\int\frac{d^2\mathbf{k}}{(2\pi)^2}
\left[E_{+}(\mathbf{k}) + 2T\ln\left(1 +
e^{-\frac{E_{+}(\mathbf{k})}{T}}\right)\right].\nonumber \\
\end{eqnarray}
The first term in the bracket is independent of $T$. This term is
removed if we redefine $F_{f}(T)$ as $F_{f}(T)-F_{f}(0)$, which
means that
\begin{eqnarray}
F_{f}(T) = -2NT\int\frac{d^2\mathbf{k}}{(2\pi)^2} \ln\left(1 +
e^{-\frac{E_{+}(\mathbf{k})}{T}}\right).
\end{eqnarray}
We then employ the transformations given by
Eqs.~(\ref{Eq:IntTransformationA})-(\ref{Eq:IntTransformationMeasure}),
and get
\begin{eqnarray}
F_{f}(T) &=& -\frac{NT}{2\pi^{2}v_{x}^{\mathrm{eff}}
v_{y}^{\mathrm{eff}}}\int_{0}^{2\pi}d\theta
\frac{1}{\left(t^{\mathrm{eff}} \cos \theta + 1\right)^{2}}
\nonumber \\
&&\times\int_{0}^{+\infty}dE E\ln\left(1 +
e^{-\frac{E}{T}}\right)\nonumber \\
&=& -\frac{3N\zeta(3)T^{3}}{4\pi v_{x}^{\mathrm{eff}}
v_{y}^{\mathrm{eff}}\left[1-\left(t^{\mathrm{eff}}\right)^{2}\right]^{3/2}},
\end{eqnarray}
where $\zeta(x)$ is Riemann zeta function. The specific heat is then
given by
\begin{eqnarray}
C_{v}(T) &=& -T\frac{\partial^2 F_{f}(T)}{\partial T^2} \nonumber \\
&=& \frac{9N\zeta(3)}{2\pi v_{x}^{\mathrm{eff}}v_{y}^{\mathrm{eff}}
\left[1-\left(t^{\mathrm{eff}}\right)^{2}\right]^{3/2}}T^{2}.
\end{eqnarray}

\subsection{Compressibility}

We next turn to compute the compressibility. To this end, we first
introduce a finite chemical potential $\mu$ into the effective
action and then re-calculate the free energy $F_{f}(T,\mu)$. After
performing calculations, we obtain
\begin{eqnarray}
F_{f}(T,\mu) &=& -NT\sum_{\omega_{n}}\int \frac{d^2
\mathbf{k}}{(2\pi)^2}\nonumber
\\
&&\times\left\{\ln\left[\left(\left(\omega_{n} - i\mu\right)^2 +
E_{+}^{2}(\mathbf{k})\right)^{\frac{1}{2}} \right]\right.\nonumber
\\
&&\left.+
\ln\left[\left(\left(\omega_{n}-i\mu\right)^2 +
E_{-}^{2}(\mathbf{k})\right)^{\frac{1}{2}}\right]\right\}\nonumber
\\
&=&-NT\sum_{\omega_{n}}\int\frac{d^2\mathbf{k}}{(2\pi)^2}
\ln\left[\left(\omega_{n} - i\mu\right)^2 +
E_{+}^{2}(\mathbf{k})\right].\nonumber \\
\end{eqnarray}
Summing over $\omega_{n}$ leads to
\begin{eqnarray}
F_{f}(T,\mu) &=& -NT\int\frac{d^2\mathbf{k}}{(2\pi)^2} \left[\ln
\left(1+e^{-\frac{E_{+}(\mathbf{k})-\mu}{T}}\right)\right.\nonumber
\\
&&\left.+\ln\left(1+e^{-\frac{E_{+}(\mathbf{k})+\mu}{T}}\right)\right],
\end{eqnarray}
where the $T$-independent term is already dropped, as what we have
done in the computation of specific heat. Again, we use the
transformations
Eq.~(\ref{Eq:IntTransformationA})-(\ref{Eq:IntTransformationMeasure})
to get
\begin{eqnarray}
F_{f}(T,\mu) &=& -\frac{NT}{4\pi^{2}v_{x}^{\mathrm{eff}}
v_{y}^{\mathrm{eff}}} \int_{0}^{2\pi}d\theta
\frac{1}{\left(t^{\mathrm{eff}}\cos(\theta) + 1\right)^{2}}\nonumber
\\
&&\times\int_{0}^{+\infty}dE E \left[\ln\left(1 +
e^{-\frac{E-\mu}{T}}\right)\right.\nonumber
\\
&&\left.+ \ln\left(1+e^{-\frac{E+\mu}{T}}\right)\right].
\end{eqnarray}
Integrating over variables $E$ and $\theta$ give rise to
\begin{eqnarray}
F_{f}(T,\mu) = \frac{NT^{3} \left[\mathrm{Li}_{3}
\left(-e^{\frac{\mu}{T}}\right) + \mathrm{Li}_{3}
\left(-e^{-\frac{\mu}{T}}\right)\right]}{2\pi v_{x}^{\mathrm{eff}}
v_{y}^{\mathrm{eff}} \left[1-\left(t^{\mathrm{eff}}\right)^{2}
\right]^{3/2}}.
\end{eqnarray}
Here, $\mathrm{Li}_{x}(y)$ is the polylogarithm function. The
compressibility can be calculated as follows
\begin{eqnarray}
\kappa(T,\mu) &=& -\frac{\partial^2 F_{f}(T,\mu)}{\partial \mu^2}
\nonumber \\
&=& \frac{NT\left[\ln\left(1+e^{\frac{\mu}{T}}\right) + \ln\left(1 +
e^{-\frac{\mu}{T}}\right)\right]}{2\pi v_{x}^{\mathrm{eff}}
v_{y}^{\mathrm{eff}} \left[1-\left(t^{\mathrm{eff}}
\right)^{2}\right]^{3/2}}.
\end{eqnarray}
In the case $\mu=0$, $\kappa$ becomes
\begin{eqnarray}
\kappa(T) = \frac{N\ln(2)T}{\pi v_{x}^{\mathrm{eff}}
v_{y}^{\mathrm{eff}}\left[1-\left(t^{\mathrm{eff}}\right)^{2}\right]^{3/2}}.
\end{eqnarray}

\section{Observable quantities at $\eta_{\psi}\neq 0$ and $z=1$ \label{App:ObservableQuantitiesSQCS} }

In order to make our paper self-contained, here we discuss the
low-energy behaviors of observable quantities in the case that Dirac
fermion acquires a finite positive anomalous dimension $\eta_{\psi}
> 0$ and the dynamical exponent remains $z=1$.

The fermion propagator is \cite{Khveshchenko01, Sachdev10}
\begin{widetext}
\begin{eqnarray}
G(i\omega_{n},\mathbf{k}) &=& \frac{1}{(i\omega_{n}\sigma_{0} -
v_{F}\mathbf{\sigma\cdot k})\left(\frac{\sqrt{\omega_{n}^2 +
v_{F}^{2}k^2}}{v_{F}\Lambda}\right)^{-\eta_{\psi}}} =
\frac{-(i\omega_{n}\sigma_{0}+v_{F}\mathbf{\sigma\cdot
k})}{(v_{F}\Lambda)^{\eta}(\omega_{n}^2 +
v_{F}^{2}k^2)^{1-\frac{\eta_{\psi}}{2}}},
\end{eqnarray}
where $v_{x}^{\mathrm{eff}}=v_{y}^{\mathrm{eff}}=v_{F}$. The
retarded propagator is
\begin{eqnarray}
G^{R}(\omega,\mathbf{k}) &=& \theta(v_{F}k-|\omega|)
\left[\mathcal{P}\frac{1}{\omega^2-v_{F}^{2}k^2} -i\pi
\mathrm{sgn}(\omega)\delta(\omega^2-v_{F}^{2}k^2)\right]
\frac{(\omega\sigma_{0} + v_{F}\mathbf{\sigma\cdot k})}{(v_{F}
\Lambda)^{\eta_{\psi}} \left(\sqrt{v_{F}^{2}k^2 -
\omega^2}\right)^{-\eta_{\psi}}} \nonumber \\
&& + \theta(|\omega|-v_{F}k) \left[\mathcal{P}
\frac{1}{\omega^2-v_{F}^{2}k^2} - i\pi
\mathrm{sgn}(\omega)\delta(\omega^2-v_{F}^{2}k^2)\right]
\frac{(\omega\sigma_{0} +v_{F}\mathbf{\sigma\cdot k})}{(v_{F}
\Lambda)^{\eta_{\psi}} \left(\sqrt{\omega^2 - v_{F}^{2}k^2}
\right)^{-\eta_{\psi}}}
\nonumber \\
&&\times \left[\cos\left(\frac{\pi\eta_{\psi}}{2}\right)
-\mathrm{sgn}(\omega)i\sin\left(\frac{\pi\eta_{\psi}}{2}\right)\right].
\end{eqnarray}
\end{widetext}
The corresponding spectral function has the form
\begin{eqnarray}
A(\omega,\mathbf{k}) &=& -\frac{1}{\pi}
\mathrm{Tr}\left[\mathrm{Im}G^{R}(\omega,\mathbf{k})\right]
\nonumber \\
&=& \frac{2}{\pi} \frac{\theta(|\omega|-v_{F}k)|\omega|
\sin\left(\frac{\pi\eta_{\psi}}{2}\right)}{(v_{F}\Lambda)^{\eta_{\psi}}
\left(\sqrt{\omega^2-v_{F}^{2}k^2}\right)^{2-\eta_{\psi}}} .
\end{eqnarray}
It is easy to get the following DOS
\begin{eqnarray}
\rho(\omega) = \frac{N}{\pi^2}\frac{1}{\eta_{\psi}}
\sin\left(\frac{\pi\eta_{\psi}}{2}\right) \frac{|\omega|^{1 +
\eta_{\psi}}}{v_{F}^{2}(v_{F}\Lambda)^{\eta_{\psi}}}.
\label{Eqn:DOSAnormalousDimension}
\end{eqnarray}
This expression clearly indicates that nonzero $\eta_{\psi}$ changes
the $\omega$-dependence of $\rho(\omega)$.

At positive $\eta_{\psi}$, the free energy becomes
\begin{eqnarray}
F_{f}(T) &=& 2NT\sum_{\omega_{n}}\int\frac{d^2\mathbf{k}}{(2\pi)^2}
\ln\left[\left(\omega_{n}^2+v_{F}^{2}k^2\right)^{\frac{1}{2} -
\frac{\eta_{\psi}}{2}}\right] \nonumber
\\
&=&\left(1-\eta_{\psi}\right)N\int\frac{d^2\mathbf{k}}{(2\pi)^2}\nonumber
\\
&&\times\left[v_{F}k-2T\ln\left(1 + e^{-\frac{v_{F}k}{T}}\right)\right].
\end{eqnarray}
It is obvious that $\eta_{\psi}$ enters only into the prefactor of
$F_f(T)$. After dropping the $T$-independent contribution, we find
that $F_{f}(T)$ is given by
\begin{eqnarray}
F_{f}(T) = -\left(1-\eta_{\psi}\right) \frac{3N\zeta(3)}{4\pi
v_{F}^{2}}T^3.
\end{eqnarray}
We get the following specific heat
\begin{eqnarray}
C_{v}(T) = \left(1-\eta_{\psi}\right)\frac{9N \zeta(3)}{2\pi
v_{F}^{2}}T^2. \label{Eqn:SpecificHeatPsi}
\end{eqnarray}
The compressibility can also be readily obtained:
\begin{eqnarray}
\kappa(T) = \left(1-\eta_{\psi}\right)\frac{N\ln(2)}{\pi
v_{F}^{2}}T. \label{Eqn:CompressibilityPsi}
\end{eqnarray}
An apparent conclusion is that, the positive $\eta_{\psi}$
\cite{Gusynin03, Zhong13, Ponte14} modifies the original linear
$\omega$-dependence of $\rho(\omega)$, but the quadratic
$T$-dependence of $C_{v}(T)$ \cite{Herbut09, Zhong13, Kaul08} and
the linear $T$-dependence of $\kappa(T)$ remain intact.

\section{RG results without the generated disorder \label{App:RGResultsDisgardingImportantTerm} }

If the important disorder-generating term
Eq.~(\ref{Eq:DisorderGenerateTerm}) is discarded naively, the RG
equations are given by
\begin{widetext}
\begin{eqnarray}
\frac{\partial v_x}{\partial \ell} &=& - \frac{1+t\lambda}{
(1-t^2)^{3/2}}\left(\Delta_0+\Delta_1+\Delta_2+\Delta_3\right)v_x +
\mathcal{H}_1(\alpha)v_x,
\\
\frac{\partial v_y}{\partial \ell} &=& -\frac{1 +
t\lambda}{(1-t^2)^{3/2}}\left(\Delta_0+\Delta_1+\Delta_2+\Delta_3\right)
v_y + \mathcal{H}_2(\alpha)v_y, \\
\frac{\partial (v_y/v_x)}{\partial \ell} &=& \left[\mathcal{H}_2
(\alpha)-\mathcal{H}_1(\alpha)\right]\frac{v_y}{v_x},
\\
\frac{d(tv_x)}{d\ell} &=& \left[-\frac{1+t\lambda}{(1-t^2)^{3/2}}
\left(\Delta_0+\Delta_1+\Delta_2+\Delta_3\right)t +
\mathcal{H}_3(\alpha)\right]v_x, \\
\frac{dt}{d\ell} &=& \mathcal{H}_3-\mathcal{H}_1(\alpha)t,
\\
\frac{d\lambda}{d\ell} &=& -\frac{1+t\lambda}{(1-t^2)^{3/2}}
\left[\left(\lambda-t\right)\left(\Delta_0+\Delta_1\right) +
\left(\lambda+t\right)\left(\Delta_2+\Delta_3\right)\right],
\\
\frac{d\alpha}{d\ell} &=& -\left[-\frac{1+t\lambda}{(1-t^2)^{3/2}}
\left(\Delta_0+\Delta_1+\Delta_2+\Delta_3\right) +
\frac12(\mathcal{H}_1(\alpha)+\mathcal{H}_2(\alpha))\right]\alpha,
\\
\frac{\partial \Delta_0}{\partial\ell} &=& \left(1-t^2\right)^{-3/2}
\left[2\Delta_0\left(\Delta_{0}+\Delta_{1}+\Delta_2+\Delta_3\right)
+4\left(1-t^2\right)\Delta_1\Delta_3+4\Delta_2\Delta_3\right]
+\left(2\lambda^2-1-t\lambda\right)f_A(\alpha)\Delta_0 \nonumber\\
&& -\left(1+t\lambda\right)^2f_B(\alpha)\Delta_0, \\
\frac{\partial \Delta_1}{\partial \ell} &=&
\left(1-t^2\right)^{-3/2}\left[2t^2\Delta_1^2+2t^2\Delta_0\Delta_1 +
4\left(1-t^2\right)\Delta_0\Delta_3 - 2t^{2}\Delta_{1}
\left(\Delta_{2}+\Delta_{3}\right)+4t^2\Delta_2\Delta_3\right]
\nonumber\\
&& -\left(1-t\lambda\right)f_A(\alpha)\Delta_1 +
\left(1+t\lambda\right)^2f_B(\alpha)\Delta_1, \\
\frac{\partial \Delta_2}{\partial \ell} &=&
\left(1-t^2\right)^{-3/2}4\left(\Delta_0+w^2\Delta_1\right)\Delta_3
- \left(1+t\lambda\right)f_A(\alpha)\Delta_2 +
\left(1+t\lambda\right)^2 f_B(\alpha)\Delta_2, \\
\frac{\partial \Delta_3}{\partial \ell} &=&
\left(1-t^2\right)^{-3/2} \left[4\left(1-t^2\right)
\Delta_0\Delta_1+4\Delta_0\Delta_2 + 4t^2\Delta_1\Delta_2 -
2\left(1-t^2\right)\left(\Delta_0-\Delta_1-\Delta_2+\Delta_3\right)
\Delta_3\right] \nonumber\\
&& +\left(1-2\lambda^2-t\lambda\right)f_A(\alpha)\Delta_3 +
\left(1+t\lambda\right)^2f_B(\alpha)\Delta_3.
\end{eqnarray}
\end{widetext}

Now suppose the system contains only RSP. We focus on the following
simplified RG equations
\begin{eqnarray}
\frac{dt}{d\ell} &=& 0, \\
\frac{d\Delta_{0}}{d\ell} &=& 2\Delta_{0}^{2}, \\
\frac{d\lambda}{d\ell} &=& \frac{\left(1+t\lambda\right)
\left(t-\lambda\right)}{\left(1+t^{2}\right)^{3/2}}\Delta_{0}.
\end{eqnarray}
It is clear that $t=t_{0}$. The solution for $\Delta_{0}$ is given
by
\begin{eqnarray}
\Delta_{0}=\frac{\Delta_{0,0}}{1-2\Delta_{0,0}\ell}
\end{eqnarray}
We find that the disorder strength $\Delta_{0}$ flows to infinity at
a finite scale $\ell_{c}=1/(2\Delta_{0,0})$. This indicates that the
system enters into a CDM phase due to RSP. The dependence of
$\lambda$ on $\ell$ is given by
\begin{eqnarray}
\lambda(\ell) = \frac{t_{0}-t_{0}\left(1-2\Delta_{0,0}\ell\right)^{
\frac{1}{2\sqrt{1+t_{0}^{2}}}}}{1+t_{0}^{2}\left(1-2\Delta_{0,0}\ell
\right)^{\frac{1}{2\sqrt{1+t_{0}^{2}}}}}.
\end{eqnarray}
It is easy to verify that $\lambda$ flows to a fixed point $\lambda
= t = t_{0}$ as $\ell\rightarrow \ell_{c} = 1/(2\Delta_{0,0})$.

If there is only $x$-RVP, the RG equations are
\begin{eqnarray}
\frac{dt}{d\ell} &=& 0, \\
\frac{d\Delta_{1}}{d\ell} &=& 2t^{2}\Delta_{1}^{2}, \\
\frac{d\lambda}{d\ell}&=&\frac{\left(1+t\lambda\right)
\left(t-\lambda\right)}{\left(1+t^{2}\right)^{3/2}}\Delta_{0}.
\end{eqnarray}
Similar to RSP, $t$ satisfies $t=t_{0}$. The solution for
$\Delta_{1}$ can be written as
\begin{eqnarray}
\Delta_{1}= \frac{\Delta_{1,0}}{1-2t_{0}^{2}\Delta_{1,0}\ell},
\end{eqnarray}
which becomes divergent at a finite scale $\ell_{c} =
1/(2t_{0}^{2}\Delta_{1,0})$. In this case, $\lambda(\ell)$ takes the
form
\begin{eqnarray}
\lambda(\ell) = \frac{t_{0}-t_{0}\left(1-2t_{0}^{2}\Delta_{1,0}\ell
\right)^{\frac{1}{2t_{0}^{2}\sqrt{1+t_{0}^{2}}}}}{1+t_{0}^{2}
\left(1-2t_{0}^{2}\Delta_{1,0}\ell\right)^{\frac{1}{2t_{0}^{2}
\sqrt{1+t_{0}^{2}}}}}.
\end{eqnarray}
Clearly, $\lambda\rightarrow t=t_{0}$ when $\ell\rightarrow\ell_{c}
= 1/(2t_{0}^{2}\Delta_{1,0})$.

Therefore, if the dynamically generated disorder given by
Eq.~(\ref{Eq:DisorderGenerateTerm}) is neglected, the disorder
strength of RSP or $x$-RVP still diverges at low energies. However,
$\lambda$ always flows to the fixed point $\lambda=t$. Accordingly,
although RSP or $x$-RVP turns the system into the CDM phase, there
is no bulk Fermi arc.

\end{document}